\documentclass[lettersize,journal]{IEEEtran}
\usepackage[utf8]{inputenc}

\usepackage{cite}
\usepackage{amsmath,amssymb,amsfonts}
\usepackage{algorithmic}
\usepackage{graphicx}
\usepackage{textcomp}
\usepackage[T1]{fontenc}
\usepackage[dvipsnames]{xcolor}
\def\BibTeX{{\rm B\kern-.05em{\sc i\kern-.025em b}\kern-.08em
    T\kern-.1667em\lower.7ex\hbox{E}\kern-.125emX}}
    
\usepackage{booktabs}
\usepackage{centernot}
\usepackage{algorithm}
\usepackage{mathtools}
\usepackage{hyperref}
\usepackage{pifont}% http://ctan.org/pkg/pifont
\usepackage{multirow}

\usepackage{tikz}
\usepackage{tikz-network}
\usepackage{tikz-3dplot}
\usepackage{tikz-cd}
\usepackage{pgfplots}
\usetikzlibrary{positioning,calc}%

\begin{document}

\title{A Survey on Large-Population Systems and Scalable Multi-Agent Reinforcement Learning }
\author{Kai Cui, Anam Tahir, Gizem Ekinci, Ahmed Elshamanhory, \\Yannick Eich, Mengguang Li and Heinz Koeppl
\thanks{This work has been co-funded by the LOEWE initiative (Hesse, Germany) within the emergenCITY center, the Distr@l-project BlueSwarm (Project 71585164) of the Hessian Ministry of Digital Strategy and Development, the German Research Foundation (DFG) as part of projects B1 and C3 of the Collaborative Research Center 1053 Multi-Mechanisms Adaptation for the Future Internet, and the State of Hesse and HOLM as part of the "Innovations in Logistics and Mobility" programme of the Hessian Ministry of Economics, Energy, Transport and Housing (HA project no.: 1010/21-12).}
\thanks{The authors are with the Department of Electrical Engineering and Information Technology, Technische Universität Darmstadt, 64287 Darmstadt, Germany. (e-mail: {\tt\small  \{kai.cui, heinz.koeppl\}@tu-darmstadt.de}).}}

\maketitle

\begin{abstract}
    The analysis and control of large-population systems is of great interest to diverse areas of research and engineering, ranging from epidemiology over robotic swarms to economics and finance. An increasingly popular and effective approach to realizing sequential decision-making in multi-agent systems is through multi-agent reinforcement learning, as it allows for an automatic and model-free analysis of highly complex systems. However, the key issue of scalability complicates the design of control and reinforcement learning algorithms particularly in systems with large populations of agents. While reinforcement learning has found resounding empirical success in many scenarios with few agents, problems with many agents quickly become intractable and necessitate special consideration. In this survey, we will shed light on current approaches to tractably understanding and analyzing large-population systems, both through multi-agent reinforcement learning and through adjacent areas of research such as mean-field games, collective intelligence, or complex network theory. These classically independent subject areas offer a variety of approaches to understanding or modeling large-population systems, which may be of great use for the formulation of tractable MARL algorithms in the future. Finally, we survey potential areas of application for large-scale control and identify fruitful future applications of learning algorithms in practical systems. We hope that our survey could provide insight and future directions to junior and senior researchers in theoretical and applied sciences alike.
\end{abstract}


\section{Introduction}
\IEEEPARstart{I}{n} recent years, the automated design of sequential decision-making through reinforcement learning (RL) \cite{sutton2018reinforcement} -- also often referred to as approximate optimal control \cite{bertsekas2019reinforcement} -- has quite prominently found application in a varied set of application areas, ranging from complex video games \cite{vinyals2017starcraft, brown2019superhuman, berner2019dota, schrittwieser2020mastering} over robotic systems \cite{kober2013reinforcement, polydoros2017survey} such as autonomous cars \cite{kiran2021deep} or unmanned aerial vehicles (UAVs) \cite{tovzivcka2018application, pham2018cooperative, chen2017socially, wang2020two, everett2021collision} to finance \cite{hu2019deep, charpentier2021reinforcement} and recommender systems \cite{afsar2021reinforcement}. While RL remains a very active research area of great importance, many important practical applications consist of more than a single agent, thus also necessitating differentiation between a variety of problem scenarios in between the fully-cooperative and fully-competitive setting. Such multi-agent problems instead fall into the subject area of so-called multi-agent reinforcement learning (MARL) \cite{zhang2021multi}. While MARL techniques sometimes perform well empirically -- typically for systems with a number of agents in the range of few to dozens \cite{muller2019generalized, schrittwieser2020mastering, papoudakis2021benchmarking} -- they suffer from a variety of difficulties. For example, the non-uniqueness of learning goals, non-stationarity of other learning agents, and concrete information structure of problems all constitute potential obstacles complicating theoretical analysis and a principled, rigorous design of MARL algorithms, see also a multitude of prior surveys on controlling and learning multi-agent systems \cite{mahajan2016decentralized, oliehoek2016concise, hernandez2019survey, zhang2021multi}. 

One of the perhaps most important practical shortcomings is the issue of scalability to large state and action spaces in the presence of many agents as well as to large numbers of agents, the latter of which will constitute the main subject of our survey. Irrespective of considering either competitive or cooperative problems, with e.g. Nash and Pareto optima as respective solution concepts, the control of multi-agent systems suffers from the so-called combinatorial nature of MARL \cite{hernandez2019survey, zhang2021multi}, sometimes also known as the curse of many agents \cite{wang2020breaking, jin2022v}. Since the joint state and action spaces of multi-agent systems grows exponentially with both the number of agents and individual agent state and action spaces in the presence of multiple agents, it has been long known that the cooperative scenario suffers from a notorious computational complexity issue \cite{bernstein2002complexity}. The competitive general Nash equilibria are similarly known to be intractable as well \cite{daskalakis2009complexity}. This issue has conferred difficulties to multi-agent control both in the theoretical understanding of MARL and in the principled design of empirically effective general learning algorithms. As a result, the complexity of general exact solutions gives motivation to a diverse spectrum of approximate or specialized algorithms, tailored to certain subclasses of multi-agent systems such as factorization approaches or limits of large-population systems discussed in this survey.

Still, large-population problems remain of significant interest to a great number of subjects in science and engineering. In contrast to prior surveys, the motivation of our survey lies instead in the learning of control for large-population systems specifically, which are surprisingly ubiquitous in nature and engineering. For examples of natural systems, see e.g. the study of epidemics \cite{kiss2017mathematics}, robotic swarm systems \cite{brambilla2013swarm}, or biological swarming such as in schools of fish \cite{couzin2002collective}, flocks of birds \cite{cavagna2014bird, perrin2021mean} and colonies of bacteria \cite{wensink2012meso}. Accordingly, we also find many use cases for large-population systems in engineering applications, for which we point to Section~\ref{sec:appl}. Generally, such systems could be understood and studied e.g. through the classical subjects of complex network theory and collective intelligence, where the focus of the former lies on decentralized systems on graphs, while for the latter the consideration is on entities interacting through simple localized interaction rules and nonetheless achieving global properties without centralized knowledge of the system state at any point \cite{hamann2018swarm}. In order to engineer and understand such large-population systems, an automated and generic design methodology via MARL is of practical interest.

In this survey, we give an account of MARL techniques and associated methods specifically designed for scalability. We begin by reviewing selected ideas from single-agent and multi-agent reinforcement learning. We then move on to present several recent promising approaches from the large-population point of view. Here, we mention and very briefly give the idea of the disjoint research areas of complex network theory, mean-field games, and collective intelligence, many ideas of which could be and have been useful for the current and future development of scalable MARL. Finally, we finish by reviewing a plentitude of applications for large-population control. Due to the large amount of literature from the multi-agent reinforcement learning community and adjacent fields, we make no claims that our survey is exhaustive and discusses all prior work. In particular, in this survey we shall focus on analysis of large-scale population systems with decision-making mostly through the lens of learning, foregoing classical control-theoretic approaches. Here, some differences between reinforcement learning and control theory include the former's focus on automated but approximate sample-based techniques, usually in discrete time, whereas multi-agent control theory instead often focuses on solving specific multi-agent problems in continuous time (e.g. formation flight \cite{lin2004local}) with theoretical guarantees under a known model class. Note however, that the intersection between techniques from both control and reinforcement learning is substantial, see also e.g. \cite{recht2019tour} for an account on reinforcement learning from the control perspective, and \cite{knorn2015overview} for a survey on control-theoretic multi-agent decision-making. In the following, we will thus refer to sequential decision-making, reinforcement learning and control interchangeably. Overall, we try to give a high-level overview of the discussed subject areas and avoid technical details, which may instead be found in the variety of excellent specialized surveys we will point towards. We hope that this allows for accessibility of the topics, and that the presentation of a variety of subjects related to large-population systems will foster the exchange of ideas between mostly disjoint fields.
\section{Sequential Decision-Making}

In this section we provide a short introductory review to single-agent and multi-agent sequential decision-making, with a focus on RL-based techniques.

\subsection{Single-agent reinforcement learning}
To begin with, sequential decision-making refers to finding a series of actions at each decision epoch in order to maximize an objective function, typically by deciding actions at each decision epoch depending on the current state of the problem (closed-loop), as opposed to deciding all actions in advance (open-loop). Therefore, in contrast to one-step decision-making as in multi-armed bandits, where only one decision is being made, this process considers the dynamics of the problem. The standard framework for modeling such problems is referred to as a Markov decision process (MDP, \cite{puterman2014markov}). MDPs consist of a state space, an action space, transition function, and a reward function. At each time step the agent / decision-maker chooses an action. Based on the action and the current state, the system moves to the next state according to the transition probabilities. Finally, for each transition the agent obtains numerical feedback via the reward function, giving rise to the maximization objective of the agent in various forms of expected cumulative rewards over future time steps. Here, the concept of an optimal policy maximizing the objective is introduced as the commonly desired solution of RL algorithms. Depending on whether e.g. an infinite time-horizon discounted objective or a finite time-horizon objective is chosen, such a policy may be a time-stationary or time-variant distribution over actions at any given state, describing an agent's choice of action at the current state \cite{puterman2014markov}.

Most of the solution methods go back to Bellman's principle of optimality \cite{bellman1966dynamic}. These dynamic programming principles introduce the concept of value functions, which describe the optimal expected future cumulative reward that may be achieved. With an appropriate choice of the objective such as the infinite-horizon discounted or finite-horizon objective, the value function will follow the so-called Bellman equation by expressing the value function in a recursive manner. Under full knowledge of the transition and reward functions, dynamic programming methods such as value iteration and policy iteration solve the Bellman equation iteratively to obtain the optimal value function, which will in turn give rise to optimal decision-making at each time and state. These methods conventionally exploit full knowledge of the model, and are referred to as optimal control methods.

In contrast, reinforcement learning aims to solve MDPs approximately and without directly accessing the transition probabilities and reward function. This is used either to solve problems with unknown models (e.g. when transition probabilities and reward function are defined implicitly through simulation), or to solve problems that are otherwise too complex to solve exactly. The goal is to learn an optimal policy by trial and error, sampling trajectories and using the corresponding reward signal from the MDP as feedback. One of the most well-known methods is the popular Q-Learning algorithm \cite{watkins1992q} which performs value iteration without accessing the model, sampling from the MDP instead. Such methods that store the value function for each state-action pair are called tabular methods. They remain limited to problems with small state and action space, as the memory needed to store the values as well as the number of state-action pairs the agent has to experience, the sample complexity, grows quickly with increasing state and action spaces. This effect is called the curse of dimensionality \cite{sutton2018reinforcement} and is typically solved by techniques of modern deep reinforcement learning.

Here, one finds two common categories of reinforcement learning through neural network approximations for tackling bigger or even continuous state and action spaces. First, value function approximation methods \cite{baird1995residual, mnih2015human} build upon tabular methods like the aforementioned Q-Learning and attempt to learn a parametrized approximation to the value function and use these to obtain a policy. Meanwhile, policy-based methods search directly over the policy space without learning the value function for maximization of the objective. Most of the latter algorithms are built on the policy gradient theorem \cite{williams1992simple, sutton1999policy}, which formulates the derivative of the expected cumulative reward with respect to the policy parameters.
In recent years, both methods have profited from the advances in deep learning \cite{lecun2015deep} and led to huge successes in agents outperforming humans in board and computer games \cite{mnih2015human, schrittwieser2020mastering}. For a more comprehensive introduction to RL, see \cite{sutton2018reinforcement}. For excellent surveys on applied RL with robotics and deep learning we refer the reader to \cite{kober2013reinforcement} and \cite{arulkumaran2017deep} respectively.

\begin{figure}[t]
    \centering
    \includegraphics[width=0.75\linewidth]{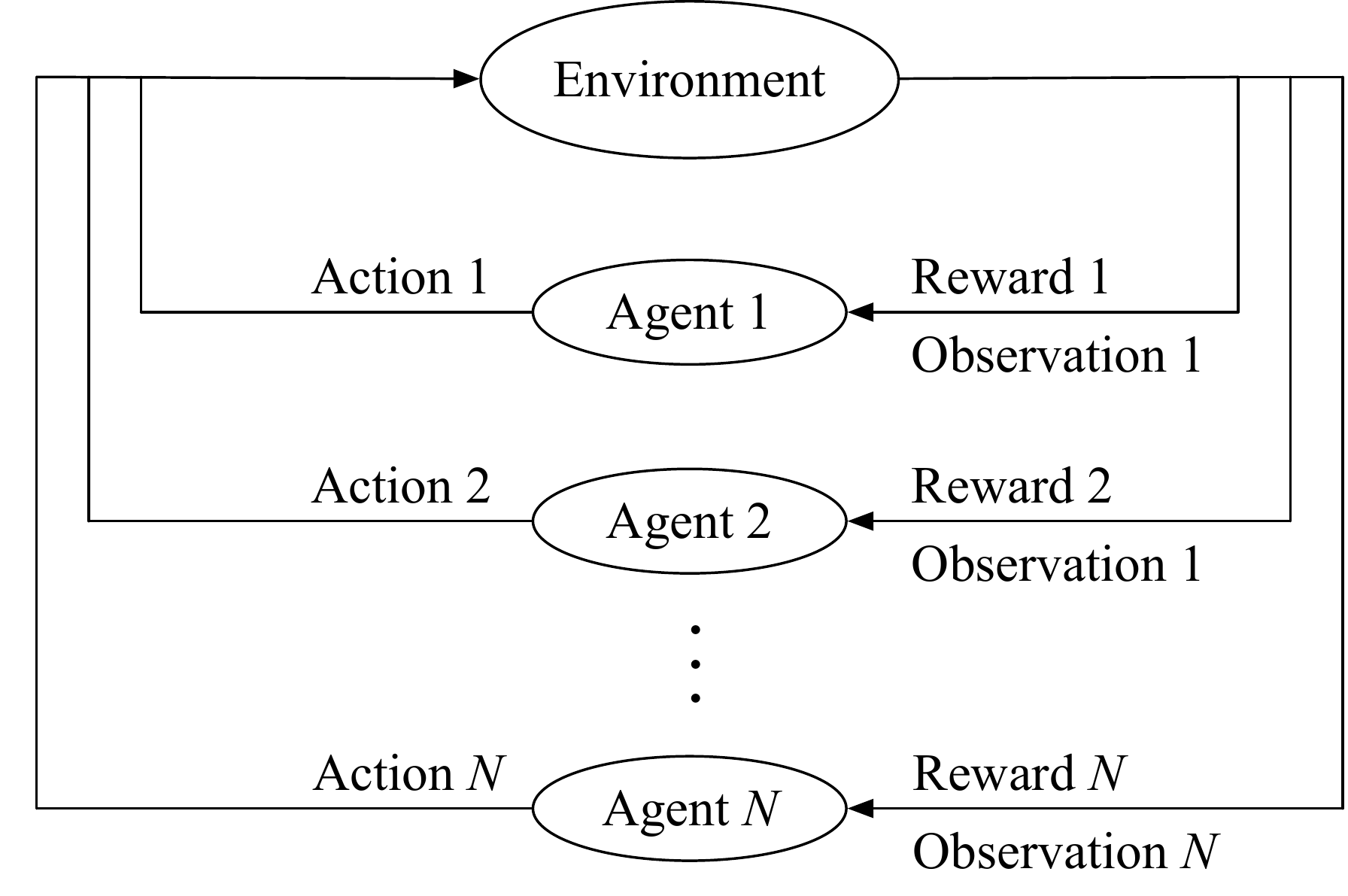}
	\caption{Visualization of a general partially-observable stochastic game with $N$ agents, each with their own observations, rewards, and actions to take.}
	\label{fig:marl}
\end{figure}

\subsection{Multi-agent reinforcement learning}
Multi-agent reinforcement learning refers instead to sequential decision-making with multiple agents. These tasks are more challenging than single-agent RL, as the transitions and rewards for each agent are now additionally influenced by other agents' actions. MARL tasks are commonly divided into the cooperative, competitive and mixed setting \cite{hernandez2019survey, zhang2021multi}. 

In the cooperative setting, the agents work together to reach a common goal. Cooperation is usually induced by a common reward function that is shared by all agents. In particular, this includes the case where each agent has its own reward function and the common goal is to maximize the average cumulative reward. As a consequence, at least for full observability, the agents can be summarily coordinated by a central decision-maker using standard single-agent RL algorithms such as by learning a joint value function for joint states and actions. These centralized models however suffer from the combinatorial nature of MARL \cite{hernandez2019survey, zhang2021multi}, or curse of many agents \cite{wang2020breaking, jin2022v}, as the overall state and action space increases exponentially with the number of agents. Apart from the combinatorial nature, partial observability in the presence of multiple agents leads to truly decentralized problems that cannot easily be solved by single-agent techniques. Building on the ideas of partially-observable Markov decision processes (POMDPs) these systems are usually modeled as Decentralized POMDPs (Dec-POMDPs) \cite{oliehoek2016concise}. It is known that complexity in general is quite high for scenarios with even just two decentralized agents \cite{bernstein2002complexity, goldman2004decentralized, amato2013decentralized}. In particular, the general problem of Dec-POMDPs has been shown to be NEXP-complete \cite{bernstein2002complexity}. Exact solutions for such systems, while possible, are therefore limited to very small systems due to complexity constraints. Here, heuristic search methods have been proposed in the past \cite{szer2012maa}. Other techniques include the reformulation of multi-agent systems as certain single-agent MDPs \cite{dibangoye2016optimally}. However, the general complexity issue remains unchanged, fundamentally motivating the search for subclasses of Dec-POMDPs that are more tractable to solve.

In the competitive setting, each agent has its own reward function and acts selfishly to maximize only its own expected cumulative reward. Typically, these tasks are modeled as zero-sum games, i.e. in order for an agent to gain, another must lose. The competition of agents is well described by the field of game theory, and accordingly the most common solution concept is that of a Nash equilibrium, i.e. a solution for all agents where no agent can gain by single-handed deviating from the solution \cite{fudenberg1991game}. Unfortunately, most of the literature is focused on the two-player or two-team to achieve tractability, which is not too interesting for large-population systems. 

Finally, that leaves the third setting of mixed competitive-cooperative scenarios, where in the most general case each agent has an arbitrary but agent-unique reward function. A common model is the partially-observable stochastic game, where each agent may have their own observations, obtain their own rewards and take actions in order to influence the system state, though there are also a variety of other useful models (cf. \cite{zhang2021multi}). See also Fig.~\ref{fig:marl} for a visualization. Again, the typical solution is the Nash equilibrium. However, similar to the aforementioned cooperative case, the computational complexity of Nash equilibria in the general case is high \cite{daskalakis2009complexity}. Exact solutions can be found as in the cooperative setting via dynamic programming \cite{hansen2004dynamic}, though similarly remaining limited to very small problem sizes due to the high complexity. 

This complexity of multi-agent control is the primary reason for the existence of a plethora of less theoretically-founded algorithms and more specialized multi-agent system models, each typically constrained to a subset of multi-agent problems that are more amenable to a computational solution while still capturing scenarios that are relevant in practice. In the following, we will begin by pointing out a subset of common learning-based MARL approaches. Following that, in the next section we will provide more specialized multi-agent system models based on graph factorizations and large-population limits. For more details on exact multi-agent control algorithms, we point to specialized surveys \cite{goldman2004decentralized, oliehoek2016concise}.

The most straightforward approach to extending reinforcement learning to a multi-agent system is by simply applying single-agent reinforcement learning methods and ignoring the multi-agent aspect of the problem. For example, under the names of parameter sharing \cite{gupta2017cooperative} and independent learning \cite{tan1993multi}, a single policy or value function is learned for all assumed homogeneous agents, and by using proven single-agent RL algorithms such as proximal policy optimization (PPO) \cite{schulman2017proximal} this has repeatedly been demonstrated to give state-of-the-art performance on a diverse set of cooperative multi-agent benchmark tasks \cite{de2020independent, yu2021surprising, papoudakis2021benchmarking, fu2022revisiting}. In particular the parameter sharing approach is one way of handling large numbers of agents tractably and even scale to unforeseen or dynamic numbers of agents in a system. However, in general convergence guarantees of approaches remain sparse, and common benchmarks remain fixated on low numbers of agents up to $10$ \cite{fu2022revisiting}, \cite[Table~2]{papoudakis2021benchmarking}. One way to understand the issue is that independent learning generates a non-stationary single-agent problem for each agent. This leads to convergence problems, especially e.g. when the number of agents is high.

A common approach to ameliorate some of the non-stationarity is the idea of centralized training with decentralized execution (CTDE), in which agents are trained offline using centralized information but executed online in a decentralized manner. Since more information is shared during training, the training phase is no longer limited by local observations of each agent, leading to potential performance improvements compared to fully decentralized training. Similar to standard RL, CTDE-based algorithms can be roughly divided into approaches based on learning a policy directly, and approaches based on learning value functions with certain structural assumptions. For learning multi-agent value functions, Value Decomposition Networks (VDN) \cite{sunehag2017value} focus on the cooperative setting where a certain joint reward function is to be maximized. They point out the phenomenon of the so-called lazy agent, where one agent is active while the other policies remain inefficient, which is resolved by learning agent-wise value functions as an additive decomposition of the joint value function. QMIX \cite{rashid2018qmix} improves upon VDN by instead allowing any non-linear monotonic combination of single agent value functions into the joint value function, allowing a richer class of value functions to be represented and improving upon state-of-the-art results in problems with up to $8$ agents. Finally, in \cite{rashid2020weighted}, two weighting functions were added to the projection operator of QMIX, further expanding the representation possibilities of the decomposed value function. On the other hand, policy-based methods such as MADDPG \cite{lowe2017multi} are a popular approach for both cooperative and competitive multi-agent systems. Here, a centralized critic improves policy gradient estimates by learning the joint value function and taking into account the actions of the other agents, which enables lower-variance policy-gradient estimates for complex agent-wise coordination. Their proposed ensemble training of policies improves the adaptability of the trained policies. COMA \cite{COMA} further adds a counterfactual baseline calculating the advantage function for each agent using a centralized critic for all agents, which efficiently addresses credit assignment problem \cite{zhang2021multi} in the multi-agent setting.

Finally, an orthogonal approach is to apply communication in the context of MARL, which may improve cooperation as part of the problem by allowing each agent to transmit relevant information to other agents. Initial works \cite{foerster2016learning, sukhbaatar16commnet} propose using independent learning to learn how to communicate while at the same time learning to cooperate in terms of actions to solve a shared task. However, the broadcasted messages that have been considered in these work could hinder the decision-making by flooding the agents with possibly irrelevant information and therefore cannot be scaled to large-scale scenarios. As a way to prevent this, TarMAC \cite{das2019tarmac} introduces a signature-based soft attention mechanism, where the messages are targeted to some agents using keys. Similarly, ATOC \cite{jiang2018learning} extends the actor-critic method by an attention unit where the senders dynamically select their collaborators from the team, and scales up to $50$ agents in both cooperative and competitive settings. \cite{wang2019learning} also proposes to limit communication by information-theoretical regularizers which minimizes the overall communication but maximizes the message information. Therefore, the agents learn to act independently most of the time and communicate when it is crucial for cooperation. 

While CTDE-based framework for solving MARL problems show promising empirical results, especially by addressing the non-stationary nature of the multi-agent systems, they may still be difficult to extend to large number of agents due to the nature of centralized critics, falling prey to the combinatorial nature of MARL. Meanwhile, independent learning, parameter sharing and communication-based techniques inherently scale to many agents, but may be difficult to train. Here, the problem of large-population MARL remains an important current and future direction. Indeed, while there exist a variety of works going towards large-population MARL such as \cite{huttenrauch2019deep, zheng2018magent, suarez2020neural}, most of them have in common an extensive manual engineering of observations, policies, and especially rewards to achieve useful results. A challenge here remains often how to solve cooperative tasks, since intuitively in the presence of many agents and a shared reward function, the reward learning signal becomes increasingly noisy due to the difficulty of credit assignment. In the next section, this survey will therefore explore a number of approaches and subject areas related to the analysis and learning of large-population systems.

\begin{table*} 
    \centering
    \caption{A selected subset of research areas and recent algorithms or modelling frameworks for multi-agent systems.}
    \label{table:taxonomy}
    \renewcommand{\arraystretch}{1.21}
    \begin{tabular}{@{}ccp{0.6\linewidth}@{}}
        \toprule 
        Category &
        Framework &
        Methodology
        \\ \midrule
        \multirow{3}{*}{RL} &
        Q-Learning \cite{watkins1992q} &
        Learns a tabular value function for optimal control with finite state and action spaces.
        \\
        &
        Value function approximation \cite{baird1995residual, mnih2015human} &
        Scales to large / continuous state spaces by learning approximated value functions.
        \\
         &
        Policy gradient methods \cite{sutton1999policy, schulman2017proximal} &
        Scales to large / continuous state and action spaces by iteratively improving a policy.
        \\ \hline
        \multirow{7}{*}{MARL} &
        Independent Learning \cite{tan1993multi}  &
        Applies single-agent RL to each agent directly, ignoring the multi-agent aspect.
        \\ &
        Parameter Sharing \cite{gupta2017cooperative}  &
        Scales to arbitrary numbers of agents by learning a single shared policy for all agents.
        \\ &
        QMIX \cite{rashid2018qmix}  &
        Decomposes and learns the joint-value as a monotonic combination of single-agent values.
        \\ &
        MADDPG \cite{lowe2017multi} &
        Policy gradient method taking into account actions of other agents in a centralized critic.
        \\ &
        COMA \cite{COMA} &
        Adds a counterfactual baseline for policy advantage estimation using a centralized critic.
        \\ &
        MAPPO \cite{yu2021surprising} &
        PPO \cite{schulman2017proximal} via independent learning, achieves state-of-the-art results \cite{de2020independent, yu2021surprising, papoudakis2021benchmarking, fu2022revisiting}.
        \\ \hline
        \multirow{10}{*}{{Graphical}} &
        Factored MDP \cite{guestrin2001, guestrin2003efficient, kok2006} &
        Local interactions are represented as a graph, factorized transition model as DBN, reward functions are additively decomposed into local reward functions.
        \\ &
        Networked Distributed POMDP \cite{nair2005, kim2006exploiting, kumar2011scalable, zhang2011coordinated} &
        Graphs represent local interactions with partial observability, transition / observation functions are factorized agent-wise, value functions are decomposed additively into local value functions.
        \\ &
        Factored Dec-POMDP \cite{oliehoek2013approximate, pajarinen2011efficient, wu2013monte} &
        Partial observability combined with coordination graphs, factorized transition model as DBN, reward function additively decomposed into local reward functions.
        \\ &
        Deep coordination graphs \cite{bohmer2020deep} &
        Hypergraphs used to represent the relationship between agents and incorporate graph convolution into the centralized training and decentralized execution structure.
        \\ &
        Hyper-Graph CoNvolution MIX \cite{bai2021value} &
        Agents determine on-the-fly hyperedges with other agents, using the signals received from the environment. 
        \\ &
        Collaborative graphical Bayesian game \cite{oliehoek2008exploiting} &
        Agents can interact with different agents at every stage of the game, resulting in a non-stationary interaction graph.
        \\ \hline
        \multirow{6}{*}{MFGs} &
        Fixed-point iteration \cite{guo2019learning, anahtarci2020q, cui2021approximately} &
        Solves contractive MFGs through application of fixed-point operator.
        \\ &
        Fictitious Play \cite{cardaliaguet2017learning, perrin2020fictitious} &
        Dampens oscillations for non-contractive MFGs, solving potential MFGs.
        \\ &
        Online Mirror Descent \cite{perolat2021scaling, lauriere2022scalable} &
        Scales and speeds up fictitious play for potential MFGs.
        \\ &
        Graphical MFGs \cite{caines2018graphon, vasal2021sequential, cui2022learning} &
        Formulates more local MFGs for agents on graphs, interacting with neighbors.
        \\ &
        MFC MDP \cite{gu2019dynamic, gu2021mean, carmona2019model} &
        Converts cooperative MARL into a high-dimensional single-agent MDP, circumventing the combinatorial nature of cooperative MARL.
        \\ \bottomrule
    \end{tabular}
\end{table*}

\section{Large-Population Systems}
While there exist a variety of important challenges in multi-agent system control such as control for a variety of possible information structures \cite{mahajan2012information} or even the question of choosing an objective -- see also the existing surveys on general multi-agent decision-making \cite{hernandez2012discrete, zhang2021multi} -- developing more specialized approaches for large-population systems seems appropriate due to the general complexity of multi-agent control \cite{bernstein2002complexity, daskalakis2009complexity}. For this purpose, we will give an overview of a diverse set of subject areas concerned with specialized large-population systems as well as their control, ranging from graph-based factorizations and complex networks theory over mean-field approximations to collective and swarm intelligence. Finally, we close with a brief taxonomy on partial observability and decentralization. For a brief overview of a selected subset of frameworks for multi-agent control, see also Table~\ref{table:taxonomy}.

\subsection{Graph-based methods} \label{sec:complex_networks}
In order to tackle the combinatorial nature of MARL and attain efficient and scalable algorithms, a popular direction is to use graphs to represent sparse interactions in-between agents, since in many practical problems, not all agents are interacting with each other. Graphical modeling approaches are highly general and allow one to model interaction between agents that is significantly more sparse than in a standard model. In the following, we will thus begin by presenting various scalable algorithms based on graphs.

\subsubsection{Factorized models}
\begin{figure}[b]
    \centering
    \includegraphics[width=0.95\linewidth]{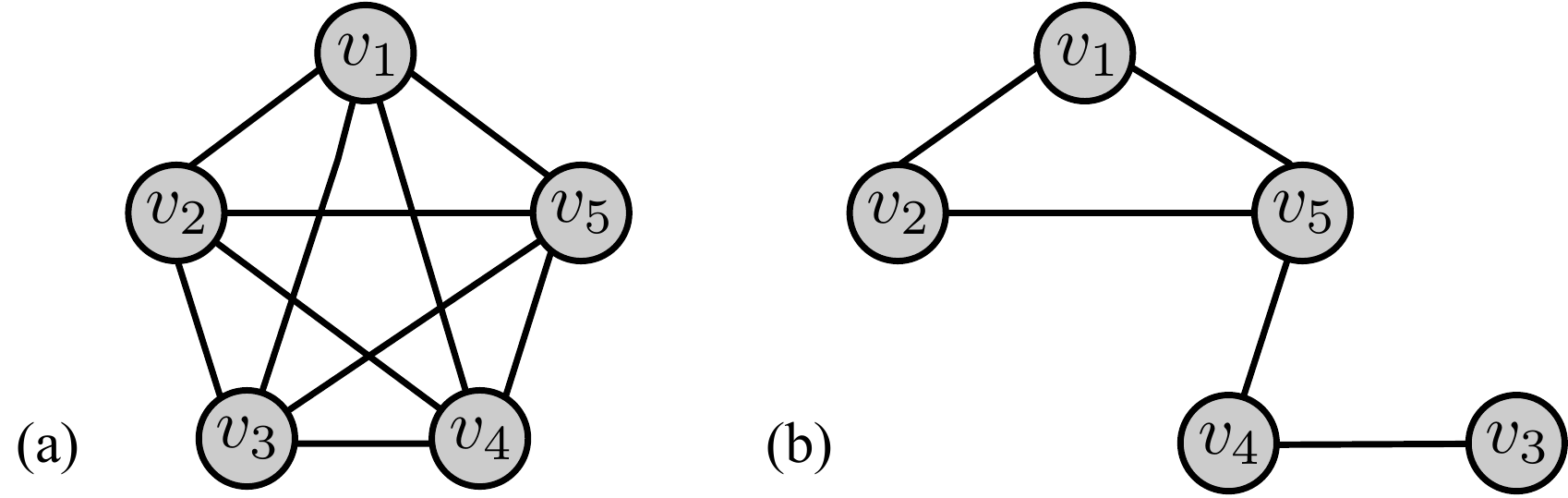}
	\caption{Visualization of (a) a fully connected graph of a system of $5$ agents labeled as $ v_i $  and (b) a coordination graph depicting local interactions in the system. The coordination graph allows factorization of the reward function into local factors and provides tractable solutions.}
	\label{fig:cg_visual}
\end{figure}

A common approach is to assume factorized graph structures in the problem. The main idea is to consider independence properties that transition, observation, and reward functions might have. Exploiting such independencies, these functions can be written in the form of smaller factors instead of functions over the high dimensional joint action and state spaces \cite{oliehoek2016concise}. For instance, consider a multi-robot navigation problem. The observations of each robot may depend only on its sensors and the local environment. Therefore the observation model can be compactly represented as a product of smaller observation functions for each robot. Factored models can be compactly represented using networked graphs where the nodes and edges show explicitly how the factors affect each other. The most straightforward way to introduce graph-based approaches to multi-agent decision-making is thus to identify and exploit such structure in the interactions of the agents a priori. In such systems, even though the behavior of all the agents affects the global reward, few agents directly interact with each other and need to coordinate. These interactions can be represented as {coordination graphs} (CG) \cite{guestrin2002coordinated}, where each agent is a node and the edges show their direct interactions. An example of a coordination graph as compared to a fully connected graph of the same population is illustrated in Fig.~\ref{fig:cg_visual}. With this graph structure, the reward model is factorized into local smaller reward functions. The interactions can also be represented using a hypergraph instead of traditional graphs when pairwise interactions are insufficient. We address these higher-order interactions separately in Section~\ref{sec:higher_order_interactions}. In the remainder of this section, we give an overview of models exploiting factorization via graph representations. 

\paragraph{Factored MDPs} Factored MDPs are one of the frameworks that employ graph-based factorization for scalability  \cite{guestrin2001, guestrin2003efficient}. In complex MDPs with large population, the global state space grows exponentially in the number of agents, therefore matrix representation of the state transition model is impractical. Factored MDPs allow tractable representations of such MDPs by factorizing the state transition model as a dynamic Bayesian network (DBN) \cite{boutilier1995exploiting}. This graph also represents the local interactions by the child-parent relations and can be interpreted as a coordination graph. Another assumption is that the reward function can be factored additively into localized reward functions which leads to more compact representations.
In factored MDPs, similar approaches as in the case of MARL are proposed for decision-making.
Authors in \cite{guestrin2001} show that the global Q-function can be additively decomposed over local functions representing the underlying graph in factored MDPs. The corresponding cost network is solved using variable elimination.
This reduces the complexity to be exponential only in the maximum number of interactions between agents instead of the number of agents.
As an alternative, the authors in \cite{kok2006} propose payoff propagation algorithm  (max-plus), that is analogous to belief propagation in Bayesian networks. They further introduce Sparse Cooperative Q-learning (SparseQ) \cite{kok2004sparse}, where the agent- or edge-based local Q-functions are updated considering its contribution to the global payoff.
\cite{guestrin2003efficient} proposes approximate linear programming and dynamic programming algorithms for factored MDPs, which are also based on variable elimination. Their dynamic programming approach shows theoretical guarantees in terms of error minimization, while the approximate linear programming algorithm is more general in terms of its assumptions. 
For large factored
MDPs, the idea of “anonymous influence” is introduced in the paper \cite{robbel2015exploiting}. In graph-based problems, they focus on the influence of variable sets rather than the identity of the variables. Exploiting this anonymity, they manage to scale up variable elimination and approximate linear programming. We note that this anonymity is a recurring theme and is also used e.g. in the formulation of mean-field models in Section~\ref{sec:mf}.
The authors in \cite{linzner2020variational} consider planning via inference for graph-based MDPs which is equivalent to factored MDP with individual agent policies. They propose an approximate method based on variational perturbation theory and attain a linear complexity in the number of agents instead of exponential, which can be scaled to larger population systems.

\paragraph{Partially-observed models} Networked Distributed POMDPs (ND-POMDPs) \cite{nair2005} are one of the frameworks that bring partial observability and coordination graphs together. 
The transition and observation functions are agent-wise factorized, and the reward function can be additively decomposed to local reward functions. In addition, it allows a set of unaffectable environment states, which cannot be manipulated by any agents. It is shown that, in this model, the value function can also be decomposed into local value functions. This decomposition also implies the {locality of interactions}, that is, an agent should not affect the policy of other agents unless they are neighbors. 
Exploiting the locality of interactions, a number of constraint optimization problem (COP) methods can be applied effectively. Authors in \cite{nair2005} propose to optimize each agent's policy with respect to its neighbors' policies in a distributed manner and propagate the optimal childrens' policies through parents in a tree-structured network.
\cite{kim2006exploiting} extends this to update the policies stochastically, allowing neighboring agents to change their policies at the same time and speeds up the convergence. Authors in \cite{kumar2011scalable} consider the planning problem as an inference problem and, using an expectation-maximization (EM) algorithm, achieve a scalable approximation for multi-agents by decomposing the global inference problem into parts involving subsets of agents.
The authors in \cite{zhang2011coordinated} extend ND-POMDP with communication between neighbors based on the coordination graph. This allows distributed learning of the joint policy, while ensuring the global convergence and scaling the learning complexity potentially linearly to the number of agents. It should be noted however, that in ND-POMDPs, the assumptions that lead to value function decomposition and therefore efficient solutions, also limit the applicability of the framework. 

Factored Dec-POMDPs constitute another framework that represents graph-based multi-agent systems and considers partial observability. Analogous to factored MDPs, the transition and observation model can be represented by a DBN and the reward function can be factored to local reward functions of smaller sets of agents. Since it does not impose agent-wise transition and observation independence as ND-POMDP, this framework is more flexible and can be applied to a wide range of problems. \cite{oliehoek2013approximate} proposes transfer planning which decomposes the problem into smaller tasks and uses the value function of those tasks as components of the factored value function of the main problem. Although this approach does not scale well with the time horizon, it shows promising results scaling up to 1000 agents. Finally, \cite{pajarinen2011efficient, wu2013monte} propose extensions to the EM algorithm for planning in factored infinite-horizon DEC-POMDPs.

\paragraph{Other recent scalable methods}
Some recent works \cite{lin2021multi, qu2020scalable, qu2020scalable2} use an underlying graph structure of agents or states and apply the correlation decay method \cite{gamarnik2013correlation} to design scalable, localized Q-functions for each agent, which only depend on the local states, actions, and rewards of the agent itself and its $k$-hop neighbors. They propose the Scalable Actor Critic algorithm (SAC) which exploits these factorizations to learn a localized policy for each agent via approximated gradients.
Another set of algorithms that make use of graph-based factorization are the distributed/consensus algorithms, which have been extensively studied to solve one-stage static optimization problems in a non-sequential manner. In these models an underlying graph is used to communicate the policy gradient estimate of each agent to the others in the network to form a consensus. We refer the reader to \cite{nedic2009distributed, varshavskaya2009efficient, agarwal2011distributed, jakovetic2011cooperative, tu2012diffusion, hong2017stochastic, nedic2017achieving, wang2019distributed} for further results in this direction.
RL for graph-based factorization has also been used for multi-task systems, where each agent has its own independent MDP but the goal is to have a joint optimal policy in the end. In other words, each agent has access to local tasks but wants to learn a global policy. In \cite{pennesi2010distributed}, the authors assume states, actions, and rewards to be local to each agent and then propose a distributed actor-critic algorithm, which is a combination of TD-step and consensus step. The authors in \cite{zhang2018fully} use a time-varying communication network to enable the agents to take local actions only based on their local information and messages received from the neighbors, such that the joint reward is maximized. Their proposed solution is to use a decentralized actor-critic framework 
with function approximation for each agent.
A similar solution is proposed in \cite{zhang2018networked} for continuous spaces, and solved using expected policy gradients \cite{ciosek2018expected}.

As we have pointed out, there is a large number of models and solution methods that take advantage of graph-based factorization, and many more we could not cover in this survey. Although these models have great theoretical potential to improve the scalability and it is acknowledged so, we note that the empirical results presented in most works still remain limited to smaller systems. Hence, we conclude that there is a great deal to be explored and investigated in this area and we imagine this would result in many new approaches and extensions to the existing ones.

\subsubsection{Complex network models} 
\label{sec:higher_order_interactions}

Apart from the diverse variety of approximate and exact control algorithms for large systems based on graph structure, the very active study of complex networks \cite{hofstad2017_RGCN} as well as dynamical processes on such complex networks \cite{albert2002statistical,barrat2008dynamical} concerns itself with understanding real-world phenomena through graph-based models with non-trivial structure. Except for mean-field approximations on large graphs -- which we will discuss also in their non-graphical version in Section~\ref{sec:mf} -- many works consider formal or rigorous analyses of a variety of graph properties and dynamical processes on graphs such as epidemics, magnetism or opinion dynamics. For example, common subject matters of complex network theory are small-world phenomena of graphs \cite{hofstad2017_RGCN}, percolation theory considering connectivity of graphs \cite{albert2002statistical} or phase transitions and synchronization in dynamical models used in statistical physics \cite{barrat2008dynamical}. Since the theory of complex networks has long been focused on understanding complex large-population systems on graphs, it seems natural that similar ideas would help in learning and controlling such large-population systems.

\begin{figure}[t]
    \centering
    \includegraphics[width=0.95\linewidth]{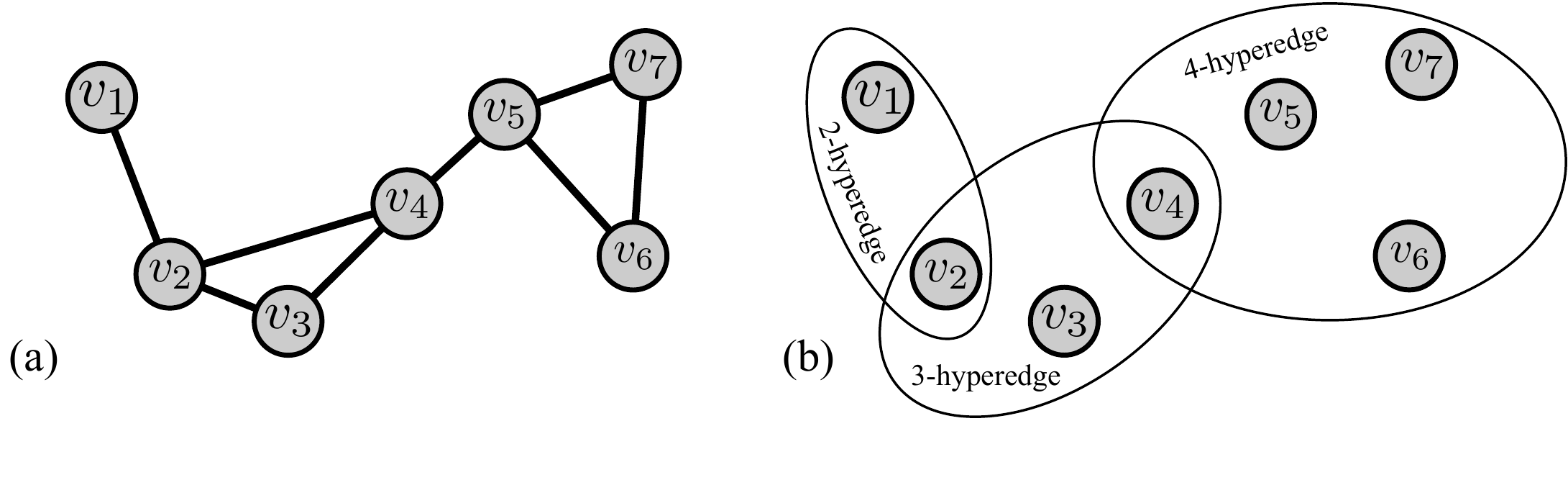}
    \caption{Comparison between a simple graph and a hypergraph, $v_i$ is the node (vertex) label, whereas circles and lines constitute the hyperedges between involved nodes. (a) Simple graph example. (b) Hypergraph example.}
    \label{fig:higher}
\end{figure}

In particular, we begin by pointing out an important emergent area of research dealing with higher-order interactions beyond simple pairwise interaction. 
In recent years there has been a great surge in interest in such higher-order dynamical models, e.g. for epidemiology \cite{bodo2016sis, landry2020effect} or opinion dynamics \cite{noonan2021dynamics, xu2022dynamics}. For example, in social networks it is natural to consider interactions that are inherently group-based and lead beyond pairwise relations. 
On the topic of higher-order interaction, there exist various excellent reviews of common modeling methods \cite{porter2020nonlinearity, battiston2020networks, battiston2021physics}, which we refer the reader to.
One of the most general ways of realizing higher-order interaction in a graphical manner is through hypergraphs, which consist of vertices and hyperedges \cite{zhou2006learning}. In contrast to standard graphs, here hyperedges are allowed to connect more than two nodes. 
See also Figure~\ref{fig:higher} for a visualization. Here, a great number of existing models from complex network theory could be explored to further improve the generality of existing frameworks and appropriately model higher-order interactive systems, which may well become increasingly important (e.g. for large-scale social networks, computer networks etc.). We thus imagine that future work could deal with higher-order interactions, be it in directed or undirected graphs, weighted or unweighted graphs and so on.

On the topic of controlled hypergraphical systems, initial works such as DGN \cite{jiang2019graph} have already made use of graph neural networks to exploit the interactions between agents. \cite{bohmer2020deep} present the framework Deep coordination graphs (DCG), which shows that using hypergraphs can further explore the representation of the relationship between agents. All these studies take the topological structure among agents into account by incorporating the idea of graph convolution into the centralized training and decentralized execution structure, and manually define the graph edges. Whereas, in Hyper-Graph CoNvolution MIX (HGCN-MIX) \cite{bai2021value}, the agent uses the environmental signals it receives to determine on-the-fly the connections, and create hypergraphs, between itself and other agents. \cite{simon2017synchronisation} also use hypergraphs to model higher-order interaction between a group of players in order to learn cooperative or competitive strategies in synchronization (symmetric hedonic) games.
A hypergraph-adapted actor-critic framework has also been proposed recently in \cite{zhang2022efficient} to employ hypergraph convolution to the centralized training with decentralized execution paradigm.
Based on the interaction structure of agents and the resulting hypergraphs they propose two novel methods for efficient information feature extraction and representation, thus leading to better collaboration policies of agents. They have used various multi-agent games to show the performance improvement of their models because of the extraction and use of higher-order interactions.
In \cite{rose2011learning}, the authors presented a multi-agent interacting system in the form of a game and then learned its equilibria for the optimal strategy. Based on the amount of information available to each agent and the communication level, the authors learned the optimal strategy of the game using numerous approaches such as fictitious play, smooth fictitious play, regret matching, reinforcement learning, etc.

\begin{figure}
    \centering
    \includegraphics[width=0.95\linewidth]{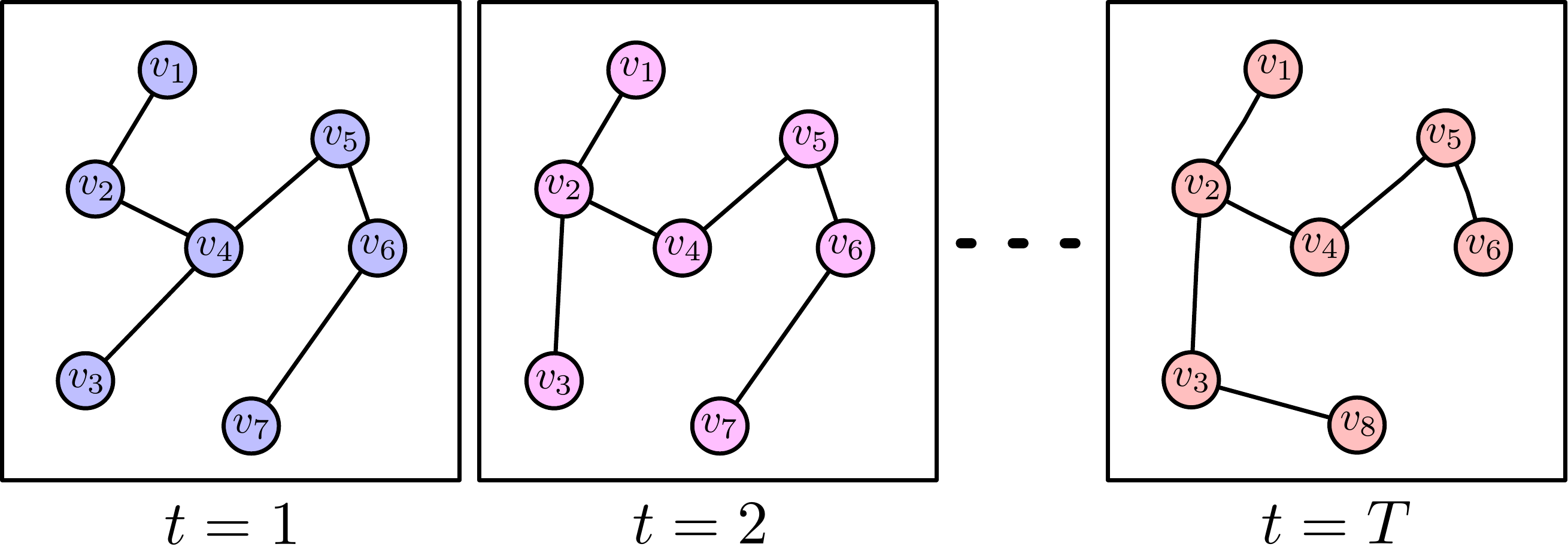}
    \caption{Visualization of an adaptive network over time. At time $t=2$, the connection (edge) between nodes $v_3$ and $v_4$ ends. Until time $t=T$, node $v_7$ leaves the network and node $v_8$ makes a connection with node $v_3$.
    }
    \label{fig:adaptive_network}
\end{figure}

Lastly, of great recent interest are also dynamical (adaptive) graph systems instead of dynamical systems on graphs, since in practice it is seldom the case that networks remain static and unchanging over time, see Fig.~\ref{fig:adaptive_network} for an exemplary visualization of both changing agents in the system and changing interactions between agents. The authors in \cite{sayama2013modeling} give a nice introduction to the concepts and properties of adaptive networks. They also emphasize its importance using applications which include but are not limited to social networks, transportation networks, neural networks, and biological networks.
In order to be more realistic and applicable to real-world scenarios, multi-agent systems not only need to be adaptive to the changing environment but also to the changing interactions, actions, and connections in the system. A recent survey \cite{nezamoddini2022survey} explains this need and highlights the existing techniques from the point of view of smart cities and Internet-of-Things.
Another survey \cite{sayed2014adaptive} presents an overview of the various distributed strategies which enable connected agents to interact locally and learn and adapt to changing traffic over the streaming network. These strategies can be non-cooperative, centralized, incremental, consensus, or diffusion based and are used for adaptation, learning, and optimization over networks.
In the factored MDPs, the assumption that the observation and transition models are fully factorized limits the applicability of the model to many real-life scenarios. A more generalized model is proposed in \cite{oliehoek2008exploiting}, under the name of collaborative graphical Bayesian games (CGBG). Here, agents may change their interactions with neighbors at every stage. This results in a non-stationary interaction graph. They propose to find a solution for this Dec-POMDP at every stage instead of finding individual optimal policies of full length. In \cite{oliehoek:uai12}, different types of agents are considered, leading to further factorization of the interaction graph through type-independence.
In \cite{creech2021dynamic} the authors present a task allocating problem in a multi-agent system with dynamic neighborhoods. They proposed four distributed strategies based on reinforcement learning, while considering that the agents have limited resources and connections with other agents may be lost or added at any time. Their solutions can be applied to any problem where the networks are dynamic and agents are self-organizing.

Especially in communication networks, using reinforcement learning to find optimal routing paths in adaptive networks is not a new direction, see e.g. \cite{mammeri2019reinforcement}. 
In \cite{boyan1993packet} the authors embedded a reinforcement learning module at every node of an adaptive switching network. The nodes use only local interactions to predict the network statistics and learn optimal routing policies to minimize the delivery time of packets under varying network loads.
Some other older and recently proposed routing algorithms include (predictive) Q-Routing based on Q learning \cite{choi1995predictive, ouvzecki2010reinforcement}, DRQ-Routing based on dual reinforcement Q learning \cite{kumar1997dual}, Hierarchical Q-Routing based on simulated annealing and hierarchy in the network \cite{lopez2011simulated}, TD-Routing based on temporal difference learning \cite{valdivia2001adaptive}, and Ants-Routing based on biological network of ants\cite{matsuo2001accelerated, subramanian1997ants, schoonderwoerd1997ant, di1998mobile}.
Mobile ad-hoc networks (MANETs) are also an example of adaptive, distributed networks. In \cite{dowling2005using} the authors use collaborative reinforcement learning to enable agents in a MANET to optimize network throughput online using feedback, without any prior knowledge or central control.  Q-routing based on Q learning has also been proposed many times to find optimal routing solutions in MANETs \cite{haraty2012manet, bitaillou2020new}. A recent comprehensive analysis of the reinforcement learning based centralized and distributed routing schemes, combined with multi-optimality routing criteria, for dynamic IoT networks is given \cite{cong2021deep}. 
Finally, adaptive networks are also a more realistic way of modeling epidemic spreading scenarios, which are frequent objects of study in complex network theory, see also \cite{Pastor-Satorras2015Complex, kiss2017mathematics}. Authors in \cite{wijayanto2018learning} have used an adaptive protection scheme using graph protection strategies, proposing n-fitted Q learning to train the model and find the optimal allocation of protection resources to maximize the number of surviving nodes. In \cite{wijayanto2019effective} also the authors study the graph protection problem by constructing minimum protective vertex covers using reinforcement learning.

Thus, higher-order and adaptive networks are a natural and realistic way to represent practical multi-agent problems. Overall, we believe that more work on the intersection of complex network theory and MARL may be a way to scale MARL algorithms to large populations with complex interactions.

\subsection{Mean-field limits} \label{sec:mf}
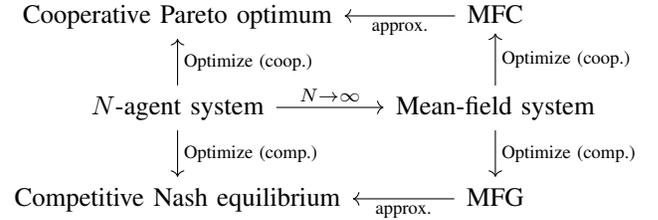
\begin{figure}
    \centering
    \[
    \begin{tikzcd}[column sep=small,ampersand replacement=\&]
    \text{Cooperative Pareto optimum} \& \arrow{l}{\text{approx.}} \text{MFC}  \\
    \text{$N$-agent system} \arrow{u}[anchor=west]{\text{Optimize (coop.)}} \arrow{d}[anchor=west]{\text{Optimize (comp.)}} \arrow{r}{N \to \infty} \& \text{Mean-field system} \arrow{u}[anchor=west]{\text{Optimize (coop.)}} \arrow{d}[anchor=west]{\text{Optimize (comp.)}} \\
    \text{Competitive Nash equilibrium} \& \arrow{l}{\text{approx.}} \text{MFG} \\
    \end{tikzcd}
    \]
    \caption{Pictorial scheme of approximation for mean-field games and mean-field control. The finite $N$-agent system is first approximated by a mean-field system, which is then solved through learning algorithms, thereby circumventing the difficult solving of the finite system. The resulting solution will be an approximately optimal solution in sufficiently large finite systems.}
    \label{fig:mf}
\end{figure}

A quite recent and related approach towards solving the scalability problem is via combining MARL with mean-field approximations and mean-field games (MFGs) \cite{yang2018mean}. Using the idea of mean-field theory (classically on graphs, see e.g. \cite{gleeson2013binary}), the many-agent problem is simplified for the infinite-agent limit.
Intuitively, in MFGs and their cooperative counterpart called mean-field control (MFC) the interaction between all agents is simplified to a two-body interaction between the mass of all agents and the behavior of any representative single agent. Most importantly, this simplification reduces the generally complex multi-agent problem into a fixed point equation (competitive) or high-dimensional single-agent problem (cooperative) with theoretical guarantees. This allows one to circumvent the otherwise difficult control problem of finite agent systems by solving the infinite agent system, which will provide an approximately optimal solution in the finite agent system, see also Fig.~\ref{fig:mf}. In particular, the cooperative MFC approach has only very recently been developed in great generality \cite{carmona2019model} and deserves further investigation, as reducing the combinatorial nature of MARL \cite{hernandez2019survey, zhang2021multi} to the better-understood curse of dimensionality of single-agent RL \cite{gosavi2009reinforcement} could allow for effective learning of scalable control.

Apart from solving limiting mean-field formulations presented in the following through RL techniques, mean-field approximations have also been applied directly to MARL. In the seminal work of \cite{yang2018mean}, the authors apply a pairwise decomposition of value functions to the competitive graph-based scenario, where value functions of each agent are assumed to depend only on the mean action of neighbors. Similar algorithms have since been proposed to extend towards partial observability \cite{subramanian2020partially} and heterogeneous agents \cite{ganapathi2020multi, subramanian2022decentralized}. In another work, combining the CTDE approach with the effect of exponential decay using k-hop neighborhoods, the authors in \cite{gu2021mean2} propose localized training - decentralized execution methods. Here the critic is not learned using global information, but using information from agents with states that are close in the sense of a network. However, so far mean-field-based approximations for MARL either remain limited to the cooperative case \cite{gu2021mean2}, or require very strong assumptions in order to obtain theoretical convergence results \cite{yang2018mean}, motivating also the solution of the following limiting mean-field systems with milder technical assumptions. 

\subsubsection{Mean-field games} Popularized by the seminal works \cite{huang2006large, lasry2007mean} in the setting of stochastic differential games, i.e. continuous-time dynamical systems described through stochastic differential equations, competitive MFGs formally assume infinitely many homogeneously modeled agents, which will allow one to handle the mass of many agents through statistical terms. Given that agents interact weakly with each other -- e.g. each agent depends continuously on the overall state distribution of all other agents -- it is often possible to obtain both theoretically rigorous approximation results for sufficiently large systems and at the same time more tractable formulations of otherwise intractable problems. Parallel to the law of large numbers, such an approach is typically asymptotically exact as the number of agents grows sufficiently large. Here, many prior works consider stochastic differential games, see also various reviews \cite{achdou2010mean, bensoussan2013mean, caines2021mean} on the very extensive body of literature on MFGs. Only recently were MFGs extended in very general form to discrete time \cite{saldi2018markov}, which allows a more accessible treatment and development of more classical learning-based algorithms, thus constituting the main focus of the review in our survey. The result is a framework for analyzing otherwise intractable competitive many-agent problems and finding approximate Nash equilibria.

In recent years, there has been an increasing focus on using RL and other learning techniques to solve such MFGs and MFC. Due to the extensive amount of literature, we will keep it brief and point towards the recent survey \cite{lauriere2022learning} for more details. Note that here, learning refers to both the classical computation of equilibria in game theory as well as modern deep RL. In contrast to finding analytic solutions, which typically remain restricted to certain special cases and require manual effort, the goal is to computationally solve arbitrary MFGs. While MFGs also find application in solving inverse RL problems \cite{yang2018learning, chen2021agent, chen2022individual} or through related but distinct mean-field approximations directly for MARL \cite{yang2018mean}, the vast majority of recent works considers algorithms for solving MFGs, which could in turn be used indirectly through model-based MARL to solve finite-sized large-population systems. The perhaps most straightforward approach is through classical fixed-point iteration \cite{huang2006large}, where one iterates over optimal policies given that all other agents play the previous policy. However, unfortunately this tends to fail in discrete-time and it is usually difficult to verify theoretical guarantees \cite{cui2021approximately}. Here, a recent approach approximates MFGs through entropy regularization, giving solvable approximate equilibria \cite{guo2019learning, guo2020general, cui2021approximately, anahtarci2021learning, guo2022entropy}. Another approach \cite{cardaliaguet2017learning, mguni2018decentralised, bonnans2021generalized} focuses on potential MFGs by adapting the classical fictitious play algorithm \cite{fudenberg1991game}. Together with modern deep techniques such as normalizing flows and deep RL \cite{perrin2021mean}, or via mirror descent \cite{perolat2021scaling, lauriere2022scalable}, such algorithms are further able at the same time to scale to large state spaces. Very recently, there have also been efforts to formulate MFGs in an optimization-based framework by using the linear program formulation of MDPs \cite{guo2022mf}. Finally, a slightly different but related scenario aims at correlated equilibria, which allow for conditioning on correlating noise between agents and have also been studied in the context of MFGs \cite{campi2022correlated}. Here, the combination of policy-space response oracles \cite{lanctot2017unified} with MFGs can similarly find a type of correlated equilibrium \cite{muller2021learning}.

\subsubsection{Mean-field control} In contrast to competitive MFGs where agents attempt to selfishly optimize their own objective, another important class of problems is described by the fully-cooperative framework of MFC \cite{andersson2011maximum, bensoussan2013mean}. In other words, the goal is to find behavior for each agent that optimizes a global objective, i.e. the solution concept shifts from Nash equilibrium to either Pareto optimum or simply the optimization of a global cost function. As a simple illustration, such a global cost function could for instance be given by the optimal foraging of food in natural swarms such as ant or bee hives, or by an optimal cell coverage in the engineering of UAV swarms for mobile cell towers. Similar to MFGs, MFC has only recently been considered in discrete time \cite{gu2021mean, carmona2019model}. Through a dynamic programming principle on an enlarged state-action space, one may reduce the MFC problem to a potentially infinite-dimensional single-agent control problem \cite{pham2018bellman, motte2019mean, gu2021mean, carmona2019model}, which will have a complexity independent of the number of agents and may be more tractable.

In the cooperative MFC problem, finding decentralized Pareto optimal policies has been the study of many recent works \cite{subramanian2019reinforcement, mondal2021approximation, carmona2019model, pasztor2021efficient}. Here, a promising approach is to formulate an equivalent high-dimensional or even infinite-dimensional MFC Markov decision process (MDP), where one finds a dynamic programming principle \cite{pham2018bellman, gu2021mean, carmona2019model} and a number of theoretically rigorous approximation properties \cite{mondal2021approximation, cui2021discrete, mondal2022can} for the MFC MDP and corresponding finite-sized large-population systems, giving the MFC approach a theoretical basis for tractable MARL. An issue that remains is that the MFC MDP typically has a function-valued action space for continuous agent state spaces, for which only limited preliminary studies have been performed \cite{pan2018reinforcement}. Here, a common idea is to discretize the continuous state space, which somewhat avoids the problem of representation and enables application of well-established tabular or approximate RL algorithms, such as Deep Deterministic Policy Gradient (DDPG, \cite{lillicrap2016continuous}) as proposed in \cite{carmona2019model}.
Alternatively \cite{gu2021mean} uses kernel regression methods to tackle the continuous state space and reduce the action space to a finite space. The authors provide an algorithm to solve the Bellman equations approximately.
Still, finding efficient reinforcement algorithms for the lifted MFC MDP remains a major challenge for solving MFC problems \cite{motte2019mean}.
For linear-quadratic mean-field scenarios, the convergence of policy gradient methods is shown by enforcing the correct form of the policy \cite{MFC_LQR}.
In a similar fashion, \cite{luo2019natural} proposed a natural actor-critic algorithm for hierarchical linear quadratic problems, while \cite{subramanian2019reinforcement, mondal2021approximation} propose policy gradient methods for stationary and general MFC problems.
In contrast to the mentioned methods, the authors of \cite{pasztor2021efficient} suggest a model-based algorithm. Similar to the single-agent case, the algorithm is sample-efficient compared to model-free approaches.

Similar ideas have been formulated also for PDE-based problems from the control community \cite{milutinovic2006modeling, zheng2021transporting}, though differences -- to name some -- include less focus on general learning-based solutions and instead on particular problems, as well as the assumption of PDE models instead of arbitrary time-discrete per-agent dynamics. While there exists some isolated prior work on learning PDE control via discretization in both time and space \cite{pan2018reinforcement}, this connection as well as a more tractable and specific design of reinforcement learning algorithms in MFC beyond discretization remains unexplored.

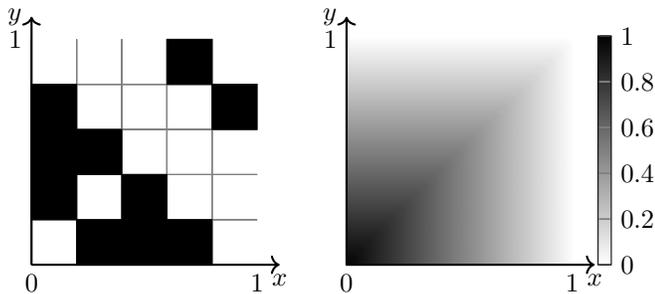
\begin{figure}
    \centering
    % \hfill
    \begin{minipage}{.42\linewidth}
        \centering
        \begin{tikzpicture}
    		[cube/.style={very thin,fill=black},
    			grid/.style={very thin,gray},
    			axis/.style={->,black,thick}]
        
        	\foreach \x in {0,0.6,...,3}
        		\foreach \y in {0,0.6,...,3}
        		{
        			\draw[grid] (\x,0) -- (\x,3);
        			\draw[grid] (0,\y) -- (3,\y);
        		}
        			
        	%draw the axes
        	\draw[axis] (0,0) -- (3.3,0) node[anchor=north]{$x$};
        	\draw[axis] (0,0) -- (0,3.3) node[anchor=east]{$y$};
        	
        	\node at (0,0) [below] {$0$};
        	\node at (3,0) [below] {$1$};
        	\node at (0,3) [left] {$1$};
        	
            \begin{scope}[shift={(0, 1.8, 0)}]
            	\draw[cube] (0,0,0) -- (0,0.6,0) -- (0.6,0.6,0) -- (0.6,0,0) -- cycle;
            \end{scope}
            \begin{scope}[shift={(1.8, 0, 0)}]
            	\draw[cube] (0,0,0) -- (0,0.6,0) -- (0.6,0.6,0) -- (0.6,0,0) -- cycle;
            \end{scope}
            
            \begin{scope}[shift={(0, 0.6, 0)}]
            	\draw[cube] (0,0,0) -- (0,0.6,0) -- (0.6,0.6,0) -- (0.6,0,0) -- cycle;
            \end{scope}
            \begin{scope}[shift={(0.6, 0, 0)}]
            	\draw[cube] (0,0,0) -- (0,0.6,0) -- (0.6,0.6,0) -- (0.6,0,0) -- cycle;
            \end{scope}
            
            \begin{scope}[shift={(1.2, 0, 0)}]
            	\draw[cube] (0,0,0) -- (0,0.6,0) -- (0.6,0.6,0) -- (0.6,0,0) -- cycle;
            \end{scope}
            \begin{scope}[shift={(0, 1.2, 0)}]
            	\draw[cube] (0,0,0) -- (0,0.6,0) -- (0.6,0.6,0) -- (0.6,0,0) -- cycle;
            \end{scope}
            
            \begin{scope}[shift={(1.2, 0.6, 0)}]
            	\draw[cube] (0,0,0) -- (0,0.6,0) -- (0.6,0.6,0) -- (0.6,0,0) -- cycle;
            \end{scope}
            \begin{scope}[shift={(0.6, 1.2, 0)}]
            	\draw[cube] (0,0,0) -- (0,0.6,0) -- (0.6,0.6,0) -- (0.6,0,0) -- cycle;
            \end{scope}
            
            \begin{scope}[shift={(1.8, 2.4, 0)}]
            	\draw[cube] (0,0,0) -- (0,0.6,0) -- (0.6,0.6,0) -- (0.6,0,0) -- cycle;
            \end{scope}
            \begin{scope}[shift={(2.4, 1.8, 0)}]
            	\draw[cube] (0,0,0) -- (0,0.6,0) -- (0.6,0.6,0) -- (0.6,0,0) -- cycle;
            \end{scope}
        
        \end{tikzpicture}
    \end{minipage}
    \hfill
    \begin{minipage}{.53\linewidth}
        \centering
        \begin{tikzpicture}
    		[cube/.style={very thin,fill=black},
    			grid/.style={very thin,gray},
    			axis/.style={->,black,thick}]
        
            \begin{axis}[scale only axis, 
                        width=86, 
                        height=86, 
                        hide axis,
                        colorbar,
                        colorbar style={
                            width=5,
                        },
                        colormap={blackwhite}{gray(0cm)=(1); gray(1cm)=(0)},
                        view={0}{90},
                        shader=interp,
                        mesh/ordering=x varies,
                        mesh/cols=3,
                        ]
        
            \addplot3[surf, 
            colormap={blackwhite}{gray(0cm)=(1); gray(1cm)=(0)},
            samples=50, samples y=30, 
            domain=0:1, domain y=0:1] 
            {1-max(x,y)};
        
            \end{axis}
            
        	%draw the axes
        	\draw[axis] (0,0) -- (3.3,0) node[anchor=north]{$x$};
        	\draw[axis] (0,0) -- (0,3.3) node[anchor=east]{$y$};
        	
        	\node at (0,0) [below] {$0$};
        	\node at (3,0) [below] {$1$};
        	\node at (0,3) [left] {$1$};
        	
        \end{tikzpicture}
    \end{minipage}
    % \hfill{}
    \caption{Visualization of the finite step graphon of a graph with $5$ nodes (left) and a limiting graphon (right). Graphons as continuous-domain versions of adjacency matrices provide a tractable way of modeling large graph limits.}
    \label{fig:graphons}
\end{figure}

\subsubsection{Graphs and partial observability} MFGs and MFC have since also been extended to interactions on graphs, with the goal of modeling sparser interactions than all-to-all. Interactions of agents should be only with a subset of other agents, which can be formulated through their neighbors on a graph, see also some early works \cite{gao2017control, caines2019graphon}. Here, graphons as limits of large graphs \cite{lovasz2012large} are used to describe MFGs on graphs, which have been formulated for static \cite{parise2019graphon, carmona2019stochastic} and dynamic agents, both in continuous and discrete time \cite{caines2018graphon, caines2019graphon, bayraktar2020graphon, cui2022learning}. See also Fig.~\ref{fig:graphons} for a visualization. The associated graphon MFGs can be considered a generalization of standard MFGs to neighbor interaction on a graph \cite{bayraktar2020graphon, bet2020weakly, cui2022learning}, or alternatively via a double limit instead as many populations of homogeneous agents, where each node in a graph represents large populations of homogeneous agents, each interacting with large populations of other neighboring populations \cite{caines2018graphon, caines2019graphon}. Here, the state of the art still remains limited to highly dense, undirected, and static graphs apart from few exceptions of higher-order \cite{cui2022motif, cui2022hypergraphon} or sparse graphs \cite{gkogkas2020graphop, lacker2020case}. 
Similarly, few works have studied partially-observed mean-field systems. Some partially-observed MFGs have been formulated in continuous time \cite{huang2006distributed, huang2014class} and with major-minor formulation which includes an external agent not part of a homogeneous swarm \cite{nourian2013mm, csen2014mean, sen2016mean, sen2019mean}. Very recently, a partially-observed system was also analyzed in discrete time \cite{saldi2019approximate}, though the partially-observed system remains restricted to deterministic mean-field. Finally, initial works have already appeared on analogous cooperative partially-observed MFC \cite{nie2022extended}. However, the development of both theory and learning algorithms for the aforementioned settings has only begun and largely remains to be explored.

\subsection{Collective swarm intelligence} \label{sec:si}
The classical field of collective intelligence has been concerned with the study of group intelligence -- as commonly observed in nature -- in form of collaborative interactions between collectives of many individual agents that behave in a decentralized manner \cite{wolpert1999introduction, wolpert2000collective}. The way of collaboration enables the swarm to accomplish tasks that are beyond the capabilities or even comprehension of their individuals. Here, the rule of interaction among the swarm considers the emergence of macroscopic intelligence arising from behavior of microscopic agents, with agent interactions that are typically highly localized and homogeneous. Due to the highly localized and homogeneous nature, such systems are scalable to almost arbitrary numbers of agents and inherently robust against failures, by redundancy of agents, as each agent can be substituted by any other. Moreover, by the fact of being decentralized, the information obtained by every agent through interaction with direct neighbors is stored locally, and not necessarily required globally. Therefore, in case of loss of certain numbers of agents, the efficiency of the whole swarm may drop but a given task is still solved to a sufficient degree. In summary, swarms are scalable to any number of agents, while maintaining their functionality without the need to define how each agent interacts individually \cite{brambilla2013swarm}.

\subsubsection{Reinforcement learning for swarm intelligence}
Given the increasing number of emerging application scenarios of collective intelligence that may be handled through the engineering of swarm behavior, e.g. in swarms of UAVs \cite{innocente2019self, su2022distributed}, synthetic biology \cite{sole2016synthetic}, or in social sciences \cite{elia2018can}, it is only natural to consider the study of collective intelligence for the control of large-population systems. In recent years, with the advance of computational power and the availability of large datasets, sophisticated deep neural network techniques have achieved state-of-the-art performances across several fields. With these advances, the field of collective intelligence aims to adapt methods of the artificial intelligence field. For instance, the fruitful intersection of collective intelligence with deep learning has achieved renewed progress in both areas, see \cite{ha2021collective} for a recent survey on this intersection. 

\begin{figure}
    \centering
    \includegraphics[width=0.65\linewidth]{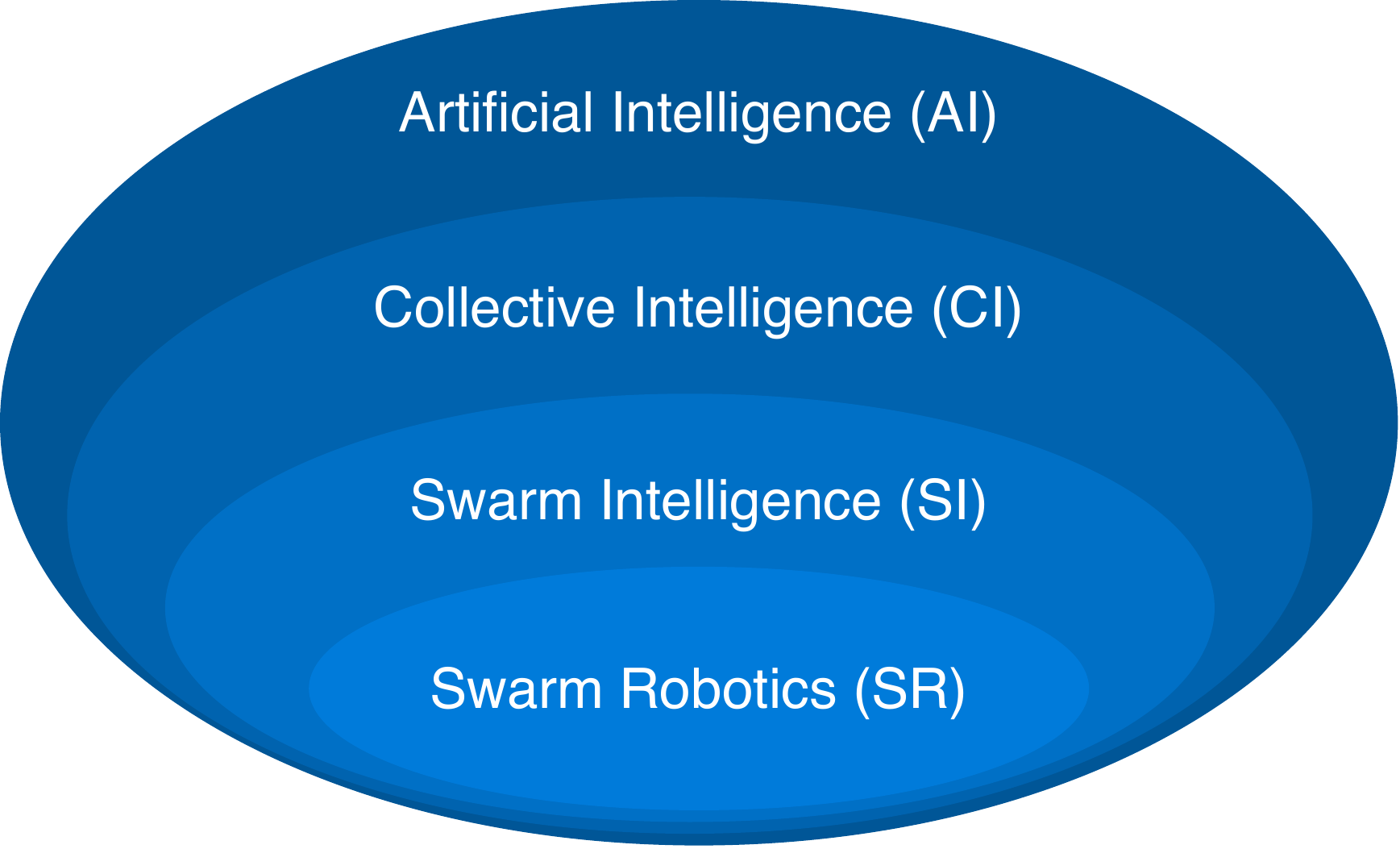}
    \caption{Taxonomy of collective intelligence. Collective intelligence can be understood as a subcategory of artificial intelligence, while swarm robotics as part of swarm intelligence are a subfield of collective intelligence.}
    \label{fig:taxonomy_CI}
\end{figure}

In connection with decision-making and reinforcement learning, the idea of self-organizing and modular, embodied agents has allowed e.g. for the design of decentralized controllers, capable of dynamically self-assembling morphologies in simulation and forming larger complex morphologies to adapt to a variety of terrains \cite{pathak2019learning}. Although there exists a variety of possibilities for the taxonomy of collective intelligence, a common classification ranging from artificial intelligence down to swarm robotics can be abstracted in Fig.~\ref{fig:taxonomy_CI}. Here, the field of swarm intelligence is considered a subfield of collective intelligence based on the collective behavior of decentralized and self-organized systems \cite{iglesias2020swarm}, while swarm robotics is based on applying the paradigms and methodologies of swarm intelligence to groups of simple homogeneous embodied agents, coordinated in a distributed and decentralized way to perform difficult swarm tasks. 

The field of MARL is mostly concerned about environment and algorithm settings directed to a simple number of agents (2-10) at the micro-level that could find a place in different application scenarios \cite{papoudakis2021benchmarking}. These scenarios are typically applications of MARL techniques on a microscopic level, which are not sufficient to deploy to large populations. In contrast, the field of collective intelligence relies on a much large number of agents involving population sizes ranging from hundreds to millions. In connection with recent advances in MARL, \cite{zheng2018magent} develops the MAgent platform to support development of many-agent reinforcement learning environments that focus on swarm intelligence applications and necessitate scalability. Through interactions among many agents, the platform facilitates learning benchmarks for the study of MARL, specifically the emergence of properties from the behavioral interaction of large populations. In turn, other works such as Neural MMO \cite{suarez2020neural} have appeared, analogously focusing on further competitive scenarios of many agents. See also discussions in the recent survey \cite{ha2021collective}. However, it seems that so far, existing works are sparse and remain limited to selfish formulations with extensive reward engineering, defining suitable per-agent rewards \cite{sovsic2018reinforcement} and per-agent observations together with appropriate policy structures \cite{huttenrauch2019deep} to achieve desirable results, see also e.g. the choice of cost function in the Battle scenario of MAgent \cite{zheng2018magent}. Here, future work could attempt to close current gaps in learning how to minimize global swarm-level objective functions in order to proceed towards engineering artificial swarm intelligence.

\subsubsection{Swarm intelligence for decision-making}
On the other hand, natural intelligent swarms such as ant or bee hives are able to solve highly complex decision-making tasks such as foraging, where under extreme decentralization the biological agents are nonetheless capable of strong coordination. In general, there are a variety of classically considered swarm intelligence tasks \cite{brambilla2013swarm, hamann2018swarm} that may not only serve as benchmarks for large-population MARL methods and building blocks for intelligent behavior, but also as a fountain of ideas through the plethora of existing prior works attempting to solve swarm intelligence through more classical or bio-inspired methods. Thus, we believe that it may be of great interest also for scalable large-population MARL to understand approaches from collective intelligence literature. In the following, we describe a subset of such tasks and how agents may organize and distribute themselves among the swarm in a swarm-intelligent manner, though for space reasons we only present the most important ideas and selected results. For more detailed information, see also the excellent surveys \cite{brambilla2013swarm, hamann2018swarm}.

In terms of spatially-organizing behaviors \cite{brambilla2013swarm}, the perhaps simplest task considered in swarm intelligence is that of {aggregation} \cite{cambier2018embodied, correll2011modeling}. In \cite{gauci2014self}, a solution for the problem of self-organized aggregation of many agents into clusters is proposed, by using only a long-range binary sensor to detect agents in their line of sight, inspired by observations of cockroaches and bees in nature studied in \cite{camazine2020self}. Results showed that the algorithm successfully aggregated at least a thousand agents into a single cluster consistently. 
Somewhat more complex, the tasks of {chain formation} and {pattern formation} consider the problem of connecting two points through agents, and deploying in a regular and repetitive pattern respectively. For the former, agents should form chains in a self-organized manner, e.g. similar to foraging ants connecting their nest with foraging areas \cite{meyer1991task}. In \cite{sperati2011self}, exploration and navigation strategies were introduced through evolutionary algorithms for effectively finding the shortest path between two points in an unknown environment. Furthermore, inspired by natural swarms, works such as \cite{fujisawa2014designing} use pheromones as swarm communicating robots without any external communication infrastructure. For pattern formation, \cite{shucker2007scalable} developed a scalable control mechanism that implements scalability, where the desired large-scale self-organizing pattern formation emerges from pair-wise interactions between agent and its nearest neighbor by following mesh abstractions called virtual springs to compute virtual forces rules of motion, i.e. repulsion or attraction such that the global behavior emerges from these simple local interactions. For further biologically-inspired methodologies in swarm intelligence, see also \cite{oh2017bio}.

Beyond relatively simple spatial organization problems, many more complex behavioral tasks are also commonly considered in swarm intelligence, including but not limited to navigation behaviors such as collective exploration and transportation, or more abstract collective decision-making such as decentralized task allocation and consensus finding. For example, task allocation assumes that challenging tasks are decomposed into a series of sub-tasks and split up among homogeneous agents \cite{lee2020task}. Consensus finding on the other hand can be found in groups of ants using pheromones to find shortest paths \cite{camazine2003self}, and may be realized through voting mechanisms \cite{montes2011majority} or specialized consensus algorithms \cite{yu2010collective}, which are an area of research in itself \cite{qin2016recent}. Similar biological inspiration is also used in collective exploration tasks, giving rise also to otherwise useful and well-known optimization techniques such as ant colony optimization \cite{dorigo2006ant} or particle swarm optimization \cite{kennedy1995particle} for navigation and global optimization respectively. Finally, in particular for embodied agents, coordinated motion, and collision avoidance are useful constituents of complex behavior, see e.g. flocking in group of birds or schooling in group of fish as impressive examples of such self-organized coordinated motion \cite{okubo1986dynamical}. Through coordinated motion, animals gain several advantages, such as a higher survival rate, more precise navigation, and reduced energy consumption \cite{parrish2002self}. Here, biological inspiration from swarm intelligence in birds and bugs has helped towards advances such as scalable collision avoidance \cite{ourari2021decentralized} and exploration \cite{mcguire2019comparative} for swarms of UAVs respectively. 

Overall, it seems that existing ideas from swarm intelligence and natural systems may help in designing appropriate algorithms for large-population systems with local interaction. More precisely, we feel that the usage of consensus algorithms as a well-studied field of research may help in the design of intelligent coordinated behavior through MARL, which has a similar flavor to communication-based methods.

\begin{table*}
    \centering
    \caption{A selected overview of partially-observable decentralized systems based on four different locality schemes.}
    \label{table:locality}
    \renewcommand{\arraystretch}{1.21}
    \begin{tabular}{cccccccc}
        \toprule 
        Title & \begin{tabular}[c]{@{}c@{}}State coupling\\ radius A\end{tabular} & \begin{tabular}[c]{@{}c@{}}Action coupling\\ radius B\end{tabular} & \begin{tabular}[c]{@{}c@{}}Observation \\ radius C\end{tabular} & \begin{tabular}[c]{@{}c@{}}Reward \\ radius D\end{tabular} & \begin{tabular}[c]{@{}c@{}}Number of\\ agents N\end{tabular} & \begin{tabular}[c]{@{}c@{}}Coop. /\\ Comp.\end{tabular} & Homog. \\ \midrule
        Factored MDPs \cite{guestrin2001} & local & local & local & local & finite & coop. & no \\
        Networked MARL \cite{zhang2018networked} & global & local & global & local & finite & coop. & no \\
        Networked Distributed POMDPs \cite{nair2005} & zero & zero & zero & local & finite & coop. & no \\
        Mean Field MARL \cite{yang2018mean, subramanian2020partially} & global & local & varies & global & finite & comp. & yes \\
        Mean Field Games \cite{lasry2007mean, huang2006large, saldi2018markov} & global & local & local & global & infinite & comp. & yes \\
        Mean Field Control \cite{gu2019dynamic, carmona2019model} & global & local & local & global & infinite & coop. & yes \\
        Mean Field Decentralized MARL \cite{gu2021mean2} & local & local & local & local & infinite & coop. & yes \\
        Scalable Actor Critic \cite{qu2020scalable, qu2020scalable2, lin2021multi} & local & local & local & varies & finite & coop. & varies \\
        \bottomrule
    \end{tabular}
\end{table*}

\subsection{Partial observability and decentralization}
Partial observability and decentralization are a key element in large-population systems, both from the theoretic and applied point of view. Without partial observability -- unless the goal is to make control design of large systems more tractable through reduction of the exponential state and action space and avoiding a global solution -- each agent must know the global state of the entire system and may thus coordinate perfectly through a global policy shared by all agents. Thus, there cannot be decentralization. On the other hand, absence of decentralization implies that agents may coordinate centrally, reducing to a centralized though potentially partially-observable single-agent problem. Meanwhile for applications, especially in large systems, it may be unrealistic or too strict of a condition to assume existence of a central planner or for all agents to know the global state of the system. Therefore, we believe that handling the different levels of partial observability and decentralization -- which we shall refer to as locality -- is of central importance and should be investigated more closely. In particular, in this survey we would like to give a new view on partial observability and decentralization by analyzing various degrees of locality that may occur in systems. We hope that this new point of view is able to identify both commonalities between existing work as well as current gaps in the theory of controlling large-population systems.

To systematically categorize and understand existing and future works on partially observed, decentralized systems, we introduce a differentiation between four different ideas of locality. This will allow us to understand relations between existing works and current limitations. We shall do this formally by identifying different `radii' of interest around an agent in terms of dynamical interactions, observations and objectives. To begin, we consider a state-coupling radius $A$, which may quantify the degree of locality in dynamical interaction, i.e. how far away do other agents have to be, in order to not affect a particular agent's transitions. For example, consider spatial epidemiological models on factored MDPs \cite{guestrin2001}, where only agents close to other infected agents may become infected by a disease. Analogously, we believe it may sometimes make sense to consider a separate action-coupling radius $B$, i.e. how far out an agent's action has the capability to affect other agent's transitions. In contrast to the first radius, it may be the case e.g. in social networks, that agents are able to choose to interact with a large quantity of agents should they wish to do so. Next, to quantify the degree of partial observability, we define an observation radius $C$ by the amount of information that an agent obtains from its neighborhood to take informed actions. This quantity is related to the locality of information structure \cite{mahajan2012information, mahajan2016decentralized} according to which actions may be performed, and it defines the level of partial observability in the system, though we stress that this is not limited to spatial radii but could also consider e.g. aggregate information or summary statistics of the whole system state such as distributions in mean-field games \cite{saldi2018markov}. Lastly, we introduce the reward radius $D$, which quantifies whether the interests of an agent lie in itself, or also in other agents. For example, in a selfish epidemiological large-population system of people, each agent may choose to care only about their own infection, while in a socially-optimal cooperative system such as in mean-field control \cite{carmona2019model}, the goal could be to minimize overall prevalence. 

Finally, we introduce a number of other, more standard categories such as the agent or behavior homogeneity, the number of agents considered in experiments, or the desired solution modalities such as competitive or cooperative. In Table~\ref{table:locality}, we give an exemplary overview of a few selected techniques discussed in the prequel, together with a rough classification of their degree of locality, though note that the taxonomy omits various specifics of approaches and it is not always straightforward to clearly classify a work. Instead, the classifications can be seen more as a rough guideline. Prior works already consider various degrees of locality, but some observations can be made. In particular, most works either attempt to handle highly localized actions or require some degree of homogeneity. Furthermore, almost all scalable approaches use graphs to decompose interactions effectively, whereas in practice such a graphical neighborhood may not always be easy to obtain or sufficiently realistic, see also the discussion in Section~\ref{sec:complex_networks}. A development of methods for missing cases may provide potential future directions.

\section{Applications} \label{sec:appl}
In practice, systems with numbers of agents too large to handle tractably through standard approaches are surprisingly ubiquitous. In the following, we give an overview over selected areas of applications that could highly profit from further research in scalable MARL methods.

\subsection{Distributed computing}
One important area of application with high accessibility (in terms of simulated data) is given by networked computers and computing applications, including for example also video games, where the advantages of MARL have been prominently and repeatedly demonstrated in scenarios with up to e.g. 10 agents \cite{vinyals2017starcraft, brown2019superhuman, berner2019dota, schrittwieser2020mastering}. On the other hand, MARL in scenarios with significantly larger population sizes such as in Neural MMO \cite{suarez2020neural} has not yet seen similar levels of success, and such benchmarks were only recently proposed. Here, future work towards a better understanding and successful design of large-scale multi-agent interaction is ongoing and could find application in making real games more interactive. We imagine that similar insights hold true also for other computing applications such as peer-to-peer systems \cite{golle2001incentives} and decentralized finance \cite{jiang2017cryptocurrency}, where automated game-theoretic analysis of user behavior could help e.g. in finding successful system designs.

Apart from very specific computing applications, another important use case of large-scale MARL may be the optimization of distributed computing itself. For example, reinforcement learning has long been used to find adaptive load balancing algorithms \cite{schaerf1994adaptive}. Even to this day, the study of load balancing remains an open problem, for example in the presence of partial observability and delayed information \cite{lipshutz2019open}, where studies on systems with large populations of servers and clients remain of importance and -- in consideration of today's increasingly large-scale computing infrastructure -- continue to be investigated using e.g. mean-field analysis \cite{gast2018refined} and learning \cite{tahir2022learning}. Some related areas of research include throughput optimization \cite{kar2020throughput}, cloud resource sharing \cite{hanif2015mean, zheng2019optimal} and edge computing \cite{banez2019mean} in systems with many devices, where mean-field approximations are often already used \cite{khudabukhsh2020generalized, hanif2015mean}. However, such formulations are typically used to analytically derive results, which has the disadvantages of requiring extensive manual efforts and considering only special cases. Here, a scalable and automated control algorithm could enable solutions for more complex and realistic models.

Scalable solutions are also a necessity in communication networks, where many agents (devices) are present. In \cite{andrews2020tracking}, RL has been proposed for a Network Inventory Manager (NIM) to learn the changing network conditions and needs of network devices. Network slicing \cite{foukas2017network} is believed to be able to provide diversified services with distinct requirements that the 5G-and-beyond systems need. Various algorithms of deep reinforcement learning have been proposed to cater to the scalability issue in the network slicing solution. We refer the reader also to \cite{wei2020dynamic,wei2020network, qi2019deep, de2019deep}. RL-based routing is also a very promising and classical direction, especially in adaptive networks with no prior information of the system, see \cite{al2015application} and Section~\ref{sec:complex_networks} for more details.

\subsection{Cyber-physical systems}
Cyber-physical systems constitute another highly important and emerging subject area. Apart from the many applications of single-agent learning in robotics \cite{kober2013reinforcement, polydoros2017survey}, more scalable MARL methods could find further applications for swarms of embodied agents. The market for unmanned aerial vehicles (UAVs) is developing rapidly, and swarms of drones could reach large-scale deployment in the near future \cite{kovalev2019analysis}, owing to their great number of potential applications. In general, swarms of terrestrial, marine and especially aerial drones could thus be a key technology for tasks such as establishing communication networks for disaster management \cite{camara2014cavalry}, performing efficient search-and-rescue missions \cite{karaca2018potential} or delivering packages \cite{shakhatreh2019unmanned}. However, deploying drone swarms in the real world is associated with a variety of coordination challenges \cite{chmaj2015distributed}, not seldom stemming from the complexity of real-world environments. Here, automated and effective decision-making for large swarms of drones remains yet an active area of research with few general design methods \cite{schranz2021swarm}.

Similarly, applications are not restricted to embodied, interactive agents and can long be found in other sectors of industry such as energy \cite{zhang2019deep}, heating \cite{castelletti2002reinforcement}, and water distribution \cite{brandi2020deep}. These wide-scale critical infrastructure sectors have similarly profited from developments in large-scale system control, see e.g. recent works on power networks \cite{bagagiolo2014mean, ma2011decentralized, ma2017charging} or
smart heating \cite{kizilkale2014collective}. In the future, critical infrastructure could profit also from achieving the effective design of more decentralized control solutions also in terms of resilience, since secure, reliable electric power and water supply remain of paramount importance to society. Here, scalable learning methods could enable key technologies such as smart grids, which may well be key to preparing and hardening the power grid against future natural and man-made disasters \cite{wang2015research}.

\subsection{Autonomous mobility and traffic control}
Many real-world scenarios associated with autonomous mobility involve many agents and therefore require scalable learning methods. Especially following the increasing connectivity of the vehicles and considering the highly dynamic and unpredictable nature of mobile systems. It constitutes one of the most challenging application areas of MARL and has received well-deserved attention from both academia and the industry in the last decades. Important challenges of the area includes safety constraints, standardized or compatible algorithms for different vehicles and integration with existing systems, e.g. manually controlled vehicles and human interaction. Here, recent work employs e.g. graph-based models \cite{troullinos2021collaborative, yu2019distributed} and distributed RL-based methods \cite{wu2010distributed} for scalability. For a more detailed view of the latest works and open challenges in autonomous mobility, we refer the reader to the recent surveys \cite{kiran2021deep, tampuu2020survey, schmidt2022introduction}.

Traffic control is another related application area that attracts attention as traffic congestion becomes more and more problematic with the increasing population and proliferation of private cars. This comes with the requirement of dealing a large number of entities such as traffic lights, vehicles and pedestrians, and therefore is an excellent potential application for MARL approaches. One way to benefit from MARL in traffic control is adaptive traffic signal control.
For example, recent work utilizes actor-critic methods combined with communication \cite{chu2019multi} and graph attention networks \cite{wang2021traffic} to help with partial observability in such problems, where both algorithms are shown to be scalable to a large real-world traffic network of Monaco city.
Another way to deal with traffic congestion through MARL is traffic routing. We refer the reader to recent works that propose mean-field based approaches to this end \cite{tanaka2020linearly, cabannes2021solving, huang2021dynamic}.

\subsection{Natural and social sciences}
Lastly, foregoing control for a moment, the study of behavior of large-scale dynamical systems is quite classical, such as through the mean-field theory originating in statistical physics for the description of magnetic materials \cite{glauber1963time}, which has since also been used as a benchmark in large-population MARL \cite{yang2018mean}. Analogous approaches are also often found in social sciences through opinion dynamics on networks of people \cite{Kashin2021Opinion,Juul2019Hipsters}, or in particular through analyzing general interacting particle systems on complex networks \cite{hofstad2017_RGCN}. Oftentimes, each agent in such models can be endowed with decision-making capabilities, leading for example to applications such as the analysis of crowd dynamics in the case of building evacuations \cite{tcheukam2016evacuation, djehiche2017evac, aurell2018mean}. 
Apart from the social sciences, one finds a variety of large-population scenarios also in the related topics of economics and finance, for which we refer to a variety of surveys \cite{hu2019deep, carmona2020applications, angiuli2021reinforcement, charpentier2021reinforcement}.

Another example of great relevance is the study of spread and control of epidemics \cite{kiss2017mathematics}, which is not restricted to biological epidemics but includes also e.g. malware spread on computer networks \cite{bruneo2012markovian, gribaudo2008analysis, VanMieghem2009Virus}, and is of interest now more than ever due to the COVID-19 pandemic. Such systems can be seen as multi-agent systems connected via complex and adaptive networks, see for example \cite{xie2020stability, tran2021optimal, abbasi2020optimal} for work in this direction. Recently, many works using RL for finding optimal decisions in epidemic situations have emerged. The works in this direction include but are not limited to \cite{liu2020microscopic, ohi2020exploring, kompella2020reinforcement, capobianco2021agent}. Here, graph-based approaches such as ones using graphons have also been used to represent heterogeneous interactions between players together with learning \cite{aurell2021finite, cui2022learning}.

\section{Future Directions}
Though there has already been some recent progress on the design of scalable MARL techniques, we imagine that future works could consider integrating ideas into MARL from a variety of recent emerging research areas.

\subsection{The limiting mean-field regime}
Of particular note to large population systems are theories of large systems and their infinite-sized limits. The theory of MFGs as well as its intersection with modern deep RL and collective intelligence remains in its infancy, both in order to engineer large-population systems, and to understand naturally occurring ones. Firstly, the development of learning algorithms for such mean-field systems remains important. Only recently have tractable learning algorithms been proposed for competitive MFGs \cite{guo2019learning, perrin2020fictitious}, though typically limited to systems fulfilling certain conditions such as monotonicity structure \cite{perrin2021generalization, perolat2021scaling} or contractivity \cite{guo2019learning, cui2021approximately}. Moreover, specifically the cooperative case of MFC has only been formulated in discrete-time very recently \cite{carmona2019model} and learning algorithms going beyond discretization-based MFC MDPs \cite{carmona2019model} or locally-optimal solutions \cite{subramanian2019reinforcement} remain yet to be developed. We imagine that further research into more tractable algorithms such as ones based on function-valued reinforcement learning \cite{pan2018reinforcement} and a combination with more classical studies of PDE system control \cite{christofides2002nonlinear} could improve the applicability of MFC-based reinforcement learning techniques to practical systems.

Secondly, the majority of MFG theories assume some sort of weak all-to-all coupling in the population, i.e. each agent has negligible but ever possibly present influence on any other agent. One advantage and primary reason is that such a construction allows one to derive mathematically rigorous results with relative ease, giving rise to well-founded approximation guarantees at least for a large subclass of problems, where weak all-to-all coupling indeed holds true. However, in order for similar MFG-based approaches to be applied to more decentralized systems without global information or locally interacting systems such as embodied robot swarms with physical collisions, it is necessary to better understand MFGs with stronger locality, both in information structure and interaction. 

To this end, work on both partially observable MFGs \cite{saldi2019approximate} and learning in e.g. graphical MFGs \cite{caines2018graphon, cui2022learning} has only yet begun and continues to remain of research interest, in particular since (i) observing the full distribution of agent states in large decentralized systems is usually unrealistic, and (ii) current graphical frameworks still assume interaction with infinitely many neighbors \cite{caines2018graphon}. For partially observable MFGs, one of the few works fusing partial observability with mean-field reinforcement learning is \cite{subramanian2020partially}, though remaining limited to very strong assumptions and not solving a limiting MFG. Meanwhile for graphical MFGs, on the one hand, less mathematically rigorous mean-field equations from physical and complex network communities such as the approximate master equations and less detailed variants \cite{gleeson2011high} could find great value in enabling scalable MARL on more realistic, bounded-degree network structures. On the other hand, rigorous approximation results are beginning to appear even for highly sparse networks, though typically requiring a variety of assumptions such as locally tree-like graphs \cite{lacker2019local} and are yet to be generalized to controlled systems. Finally, the respective cooperative counterparts of partially observable and graphical MFGs remain to be explored.

\subsection{Higher-order complex networks}
Analogously, we believe that further work combining the worlds of complex network sciences and graphical reinforcement learning may be of great interest, as there have been great recent efforts in extending complex network theories to higher-order or adaptive networks. Apart from applying graphical mean-field approximations as mentioned in the prequel, recent developments in complex network theory consider more than simple locally tree-like graphs and allow for clustering or higher-order interactions, e.g. in the form of hypergraphs.
Hypergraphs have already started gaining attention since it has been shown that simple pairwise interactions are insufficient for representing complex, real systems \cite{battiston2020networks}. For example, \cite{iacopini2019simplicial} contribute to understanding the higher-order interactions in complex systems using simplicial models and their relation to hypergraphs. They also highlighted that pair-wise interactions are insufficient to accurately represent complex social networks in which contagion occurs due to interactions in different group sizes. While some intersecting work between hypergraphs and MARL has been mentioned, we believe that further exploration in this direction may be fruitful.

Additionally, modeling approaches for representing large adaptive networks have emerged very recently and could be applied in a control setting. This is important, since while many real-life processes can be modeled through graphical interactions, such interactions usually change over time. For example, recently, \cite{garbe2022flip} use a general idea of flip processes to create time dependent trajectories in the space of graphons and then study large graphs from the dynamical systems perspective. Here, further work could integrate control elements with large adaptive graphs. Similarly, another approach could be to continue research on dynamical processes on random geometric graphs, which have been used e.g. to study epidemics \cite{preciado2009spectral} or synchronization of oscillators on graphs \cite{diaz2009synchronization} and allow for the linking of graph connections to an underlying spatial structure.
To obtain an understanding of the theory of random geometric graphs, which could be used to study the relations between spatial agents in a multi-agent system, we refer the reader to \cite{penrose2003random, walters2011random}.
We believe that the use of MARL to learn optimal policies in such scenarios is a very fruitful direction and should be further investigated, perhaps also by considering the recently emerging graph convolutional networks \cite{wu2020comprehensive} for reinforcement learning \cite{jiang2019graph} together with limiting graphon-based formulations \cite{ruiz2020graphon}. Overall, a combination of graphical modeling schemes with multi-agent decision-making seems like a promising direction.

\subsection{Intersectional and application-oriented work}
Lastly, at the intersection of swarm intelligence and MARL lies another promising approach to tractably engineering artificial swarm intelligence for systems with many interacting agents. Work at this intersection has only begun and we imagine that further work could lead to great advances in both engineering and understanding swarm behavior, in particular since so far, there appears to be no definitive toolchain or framework for computationally designing intelligent swarm systems \cite{schranz2021swarm}. Here, we imagine that learning-based approaches utilizing a number of powerful ideas from MARL, MFGs, and complex graphical networks could provide such a toolchain in the near future, or supplement existing ideas \cite{huttenrauch2019deep}. 

In particular, we imagine that automated design tools for swarm robotics could enable a plethora of interesting applications for which current methods would necessitate an extensive manual design of algorithms. For example, many of the difficult and abstract subtasks in the study of swarm intelligence mentioned in Section~\ref{sec:si} and \cite{hamann2018swarm, brambilla2013swarm} could likely be handled by recent advances in scalable MARL. Similarly, we believe that an application of approaches based on limit theories could propel forward many aforementioned fields of research, ranging from computing over cyber-physical and embodied swarm systems to natural and social sciences.

\section{Discussion}
The primary challenge we have discussed in detail is scalable analysis and understanding of large-population systems in the context of MARL for automated design of sequential decision-making. Indeed, in practice we often encounter systems that are too large to be tractably handled by standard control and reinforcement learning techniques, as evidenced also by the large amount of applications and ongoing research in the field. While there has been some recent progress on the issue of scalable learning in large-population systems, important open research questions and synergies with existing fields of research remain that could be addressed in future works. In particular, we have given an overview of a variety of long-standing subject areas dealing with large-scale systems such as mean-field games, collective intelligence, and complex networks. It is our belief that intersectional works will be crucial to the further development of tractable MARL. We hope that this survey proves useful in bridging the diverse set of research subjects on large-scale multi-agent systems and providing new perspectives for future works.

\bibliographystyle{IEEEtran}
\bibliography{references}

\end{document}